\documentclass[lettersize,journal]{IEEEtran}
\usepackage{microtype}
\usepackage{graphicx}
\usepackage{booktabs} %

\usepackage{amsmath, amsfonts}
\usepackage{amsthm}
\usepackage{bm}
\usepackage{color, colortbl}
\usepackage{enumitem}
\usepackage{amssymb}
\usepackage{mathtools}
\usepackage[font=small]{caption}
\usepackage{subcaption}
\usepackage[normalem]{ulem}
\usepackage[hyphens]{url}
\usepackage[scientific-notation=true]{siunitx}

\newtheorem{lemma}{Lemma}
\newtheorem{theorem}{Theorem}
\newtheorem{corollary}{Corollary}

\newtheorem{assumption}{Assumption}

\usepackage{mwe}
\usepackage{graphicx,caption}
\graphicspath{ {../} }
\usepackage{empheq}
\usepackage{algorithm}
\usepackage{algpseudocode}
\usepackage{thmtools}
\usepackage{thm-restate}

\algnewcommand{\LineComment}[1]{\State \(\triangleright\) #1}

\DeclareMathOperator{\x}{\textbf{x}}
\DeclareMathOperator{\y}{\textbf{y}}

\DeclareMathOperator{\thetab}{\bm\theta}

\DeclareMathOperator{\B}{\mathcal{B}}

\DeclareMathOperator{\X}{\textbf{X}}

\DeclareMathOperator{\Finf}{F^{\mathrm{inf}}}
\DeclareMathOperator{\G}{\textbf{G}}

\DeclareMathSymbol{\shortminus}{\mathbin}{AMSa}{"39}

\newcommand{\lrVert}[1]{\left\Vert #1 \right\Vert}
\newcommand{\Ebatch}[1]{\mathbb{E}^r \left[#1\right]}
\newcommand{\Etot}[1]{\mathbb{E} \left[#1\right]}

\newcommand{\rev}[1]{{#1}}

\newcommand{\Hm}{\mathcal{K}}

\algblock{ParFor}{EndParFor}
\algnewcommand\algorithmicparfor{\textbf{parallel for}}
\algnewcommand\algorithmicpardo{\textbf{do}}
\algnewcommand\algorithmicendparfor{\textbf{end\ parallel for}}
\algrenewtext{ParFor}[1]{\algorithmicparfor\ #1\ \algorithmicpardo}
\algrenewtext{EndParFor}{\algorithmicendparfor}

\algblock{Execute}{EndExecute}
\algnewcommand\algorithmicex{\textbf{execute}}
\algrenewtext{Execute}[1]{\algorithmicex\ #1}
\algrenewtext{EndExecute}{\textbf{end\ execute}}
\newcommand{\subhead}[1]{\multicolumn{1}{c}{#1}}%
\usepackage{multirow}

\usepackage{cite}
\allowdisplaybreaks

\begin{document}

\title{Flexible Vertical Federated Learning with Heterogeneous Parties}
\author{Timothy Castiglia, Shiqiang Wang, and Stacy Patterson 
    \thanks{T. Castiglia and S. Patterson are with the Department of Computer Science,
        Rensselaer Polytechnic Institute, 110 8th St, Troy, NY 12180 USA,
        {\tt\small castit@rpi.edu, sep@cs.rpi.edu}.}%
    \thanks{
        S. Wang is with the IBM Thomas
        J. Watson Research Center, Yorktown Heights, NY 10598 USA,
        {\tt\small wangshiq@us.ibm.com}.}%
}


\maketitle

\begin{abstract}
    We propose Flexible Vertical Federated Learning (Flex-VFL), a distributed machine algorithm that trains a smooth, non-convex function in a distributed system with vertically partitioned data. We consider a system with several parties that wish to collaboratively learn a global function. Each party holds a local dataset; the datasets have different features but share the same sample ID space. The parties are heterogeneous in nature: the parties' operating speeds, local model architectures, and optimizers may be different from one another and, further, they may change over time. To train a global model in such a system, Flex-VFL utilizes a form of parallel block coordinate descent, where parties train a partition of the global model via stochastic coordinate descent. We provide theoretical convergence analysis for Flex-VFL and show that the convergence rate is constrained by the party speeds and local optimizer parameters. We apply this analysis and extend our algorithm to adapt party learning rates in response to changing speeds and local optimizer parameters. Finally, we compare the convergence time of Flex-VFL against synchronous and asynchronous VFL algorithms, as well as illustrate the effectiveness of our adaptive extension. 
\end{abstract}

\section{Introduction} \label{intro.sec}

In modern distributed systems, data are often generated by multiple parties 
and must remain on premises to follow regulations 
(e.g., GDPR~\cite{voigt2017eu}, HIPAA~\cite{hipaa})
and protect sensitive personal information.
Federated learning algorithms~\cite{pmlr-v54-mcmahan17a, kairouz2019advances, DBLP:journals/tist/YangLCT19} were introduced
to provide methods for training machine learning 
models in distributed systems without the need to share raw data between parties. 
In these algorithms, data-owning parties train models locally
and share intermediate information with a parameter server
to update the global model.
Federated learning has many important applications including 
personalized healthcare, 
smart transportation, 
and predictive energy systems~\cite{kairouz2019advances, zhou2021survey}.

\begin{figure}[t]
    \centering
    \includegraphics[width=0.4\textwidth]{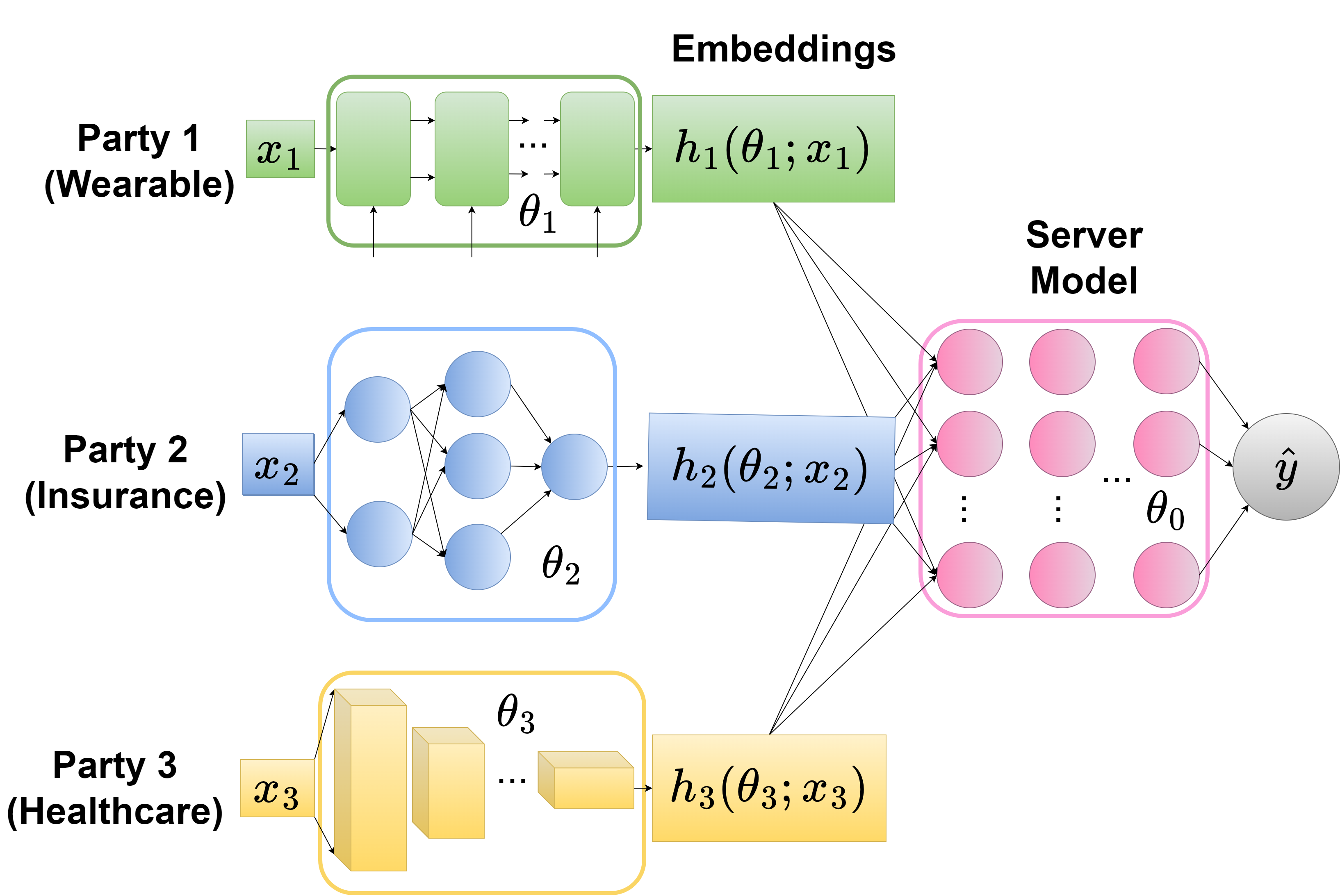}
    \caption{Example global system model $\bm\Theta$, 
    where party $1$'s model $\thetab_1$ is an LSTM,
    party $2$'s model $\thetab_2$ is an MLP,
    party $3$'s model $\thetab_3$ is a CNN,
    and the server model $\thetab_0$ is deep neural network.
    }
    \label{vflmodel.fig}
\end{figure}

\emph{Vertical Federated Learning} (VFL) is an important class of federated learning. 
In VFL,
parties' local datasets share a common sample ID space but
have different feature sets~\cite{DBLP:journals/tist/YangLCT19, liu2019communication, castiglia2022compressed}.
\rev{This is in contrast to \emph{Horizontal Federated Learning} (HFL), where 
all parties' datasets share the same feature space, but each party's data 
corresponds to a distinct set of sample IDs~\cite{
    pmlr-v54-mcmahan17a, ZhouC18Kstep, SattlerWMS20, ChenSJ20, jamali2022federated}.}
For example, consider a case where a healthcare provider, insurance company, and 
wearable device manufacturer wish 
to collaboratively train a model to identify diseases 
without directly sharing raw user information with one another~\cite{Sun2019-ki}.
These parties store information about the same people, 
but each party has a distinct set of information for each individual. 
In VFL, each party typically trains a local feature extractor, while
a central server trains a fusion model.
The parties periodically exchange intermediate information
for updating their local models.
We provide an example of a VFL model setup in Figure~\ref{vflmodel.fig}.

Many VFL algorithms
assume parties have identical local training: 
each party uses the same local optimizer to update its model, 
typically standard SGD~\cite{liu2019communication, chen2020vafl, 9415026, castiglia2022compressed, FDML, Fu2022VFLcache, ZhangGQZWLZ22}.
However, in practice, parties store different types of feature sets ranging from medical images to income values. 
Centralized machine learning accommodates such multimodal data 
by supporting different local feature extractors and optimizers
for each feature set~\cite{HazarikaZP20, HanCP21MOSEI, LinG0ZHRH21, narkhede2021gas}.
However, such heterogeneous local optimizers have not been accommodated in previous VFL works.

Additionally, it is typically assumed that each party takes the same amount of time to update its local model.
The complexity of a party's model and choice of local optimizer can affect its local training time.
There can also be differences in party operating rates due to 
differing compute capabilities
or a varying workload in each party due to colocated jobs. 
Further, a party's operating rate may change over time.
The slowest parties in the system, the stragglers, 
can become a bottleneck in the convergence of VFL 
algorithms~\cite{ReisizadehTMHP19,Xu2019stragglers,Reisizadeh2020stragglers}.

Previous VFL works have addressed some of these issues.
Asynchronous VFL algorithms~\cite{chen2020vafl,FDML, ZhangGDH21,Shi2022Async} were proposed as a means
to avoid the straggler bottleneck. 
These algorithms allow parties to train at their own pace, and 
send updates whenever they are able.
However, training with stale information for long periods has been shown to degrade overall
model performance, overwhelming the benefits of flexibility 
to heterogeneous party operating speeds~\cite{CuiZGGX16, dai2018learning}.
Some VFL algorithms using common optimizers other than standard SGD have been 
proposed~\cite{mugunthan2021MultiVFL, yang2021model, gu2021privacy, ZhangGDGBPH21, xie2022VFLADMM}, 
but they require that all parties apply the same local optimizer for training.
No previous VFL works incoporate both the effect of 
time-varying operating speeds and heterogeneous local optimizers.

\rev{In this work, we consider a heterogeneous VFL setting where parties
have different, time-varying, operating rates and different local optimizers.
We seek to answer the following questions.}
\textit{Is there a method of VFL that can be flexible 
to the inherently asynchronous nature of distributed parties?}
\textit{Can we analyze the effect of heterogeneous 
party local optimizers in a VFL setting?}
And finally, \textit{can we generalize the analysis to gain insight into
the effect of commonly used local optimizers on VFL convergence?}

To address these questions, 
we propose Flexible Vertical Federated Learning (Flex-VFL).
Flex-VFL is a communication-efficient distributed learning algorithm for vertically partitioned data
that is robust to heterogeneous party operating speeds, 
lets parties utilize different local feature extractors and optimizers,
and supports a variety of local optimizers. 
In Flex-VFL, rather than the server waiting for all parties to complete a
given number of local iterations, each party completes as many iterations 
as possible within a specified timeout.
The parties synchronize with the server after this set amount of time
has passed. 
\rev{Our approach serves as a middle-ground between fully synchronous and asynchronous
methods: training is not slowed down by stragglers,
but parties still synchronize regularly to avoid training with stale information for long periods of time.
Further, unlike previously proposed VFL algorithms, Flex-VFL allows parties to customize their local
optimizers based on their local data and feature extractor architecture.
}

Flex-VFL is the first theoretically-verified VFL algorithm that 
has convergence guarantees when parties use different local optimizers.
To represent these optimizers in our analysis, we apply arbitrary weights to 
party gradients at each local iteration.
This approach was first introduced by Wang et al.~\cite{Vincent_Poor2021},
but it has never been applied to the VFL setting or generalized for 
time-varying optimizer parameters.
Our convergence results are generalizable to many variations of VFL,
and provides novel insights into the benefits and drawbacks of
local momentum~\cite{Vincent_Poor2021},
proximal terms~\cite{li2018federated}, 
and variable learning rates 
on the convergence of VFL algorithms.
Specifically, we show that these optimizers improve convergence speed 
but may increase the error at convergence.
\rev{Our analysis shows that proper tuning of party learning rates
can help offset the error introduced by party heterogeneity.}

Our work also provides an adaptive extension to Flex-VFL known as Adaptive Flex-VFL:
a meta-optimization algorithm that improves the convergence rate.
In systems where other jobs may be colocated on the participating devices, 
party operating speeds may vary over the course of training.
In these cases, it is a challenge to choose hyperparameters to accommodate such
heterogeneity over time. 
Based on our theoretical results on convergence in Flex-VFL,
the server can gather party information in each global round 
to optimize the party learning rates.
Adaptive Flex-VFL is designed to be robust to 
heterogeneous and time-varying party speeds and optimizer parameters.

Our main contributions are as follows: 
\begin{enumerate}[leftmargin=*]
    \item We propose Flex-VFL, a Vertical Federated Learning algorithm that is robust to 
        heterogeneous, time-varying parties. Flex-VFL supports a large class of SGD variants such
        as SGD with local momentum, proximal steps, and variable learning rates.
    \item \rev{We provide convergence analysis for Flex-VFL and show that the error incurred by heterogeneous parties can be offset with the proper choice of learning rates in each round.}
    \item \rev{We propose Adaptive Flex-VFL, an extension of Flex-VFL where 
    each party's learning rate is tailored to its speed and optimizer parameters
        at each round.}
    \item \rev{We experimentally compare Flex-VFL with other VFL algorithms using
    both simulated and real-world party operating speeds.
    We find that Flex-VFL outperforms
purely synchronous and asynchronous VFL algorithms, reaching target accuracy
        up to $4\times$ as fast.} 
    We also compare Adaptive Flex-VFL and Flex-VFL using 
        real-world time-varying party operating speeds, 
    and show up to a $30\%$ time-to-target improvement. 
\end{enumerate}

The rest of the paper is structured as follows. In Section~\ref{related.sec}
we discuss related work. Section~\ref{problem.sec} introduces our system model and problem
formulation. We present Flex-VFL in Section~\ref{alg.sec}. 
We analyze the convergence of our algorithm in Section~\ref{adapt.sec},
and present an
optimization method to improve convergence speed by 
adapting party learning rates in Section~\ref{adapt2.sec}. 
In Section~\ref{exp.sec} we present our experiments. Finally,
we conclude in Section~\ref{conclusion.sec}.

\section{Related Work} \label{related.sec}

Many works in HFL tackle
the challenges of high-latency communication with the use
of local iterations. 
\rev{These works analyze the effect of local iterations on convergence~\cite{ZhouC18Kstep, koloskova2020unified, jamali2022federated}.} 
Castiglia et al.~\cite{castiglia2021multilevel} and Wang et al.~\cite{Vincent_Poor2021}
both propose HFL algorithms that support heterogeneous party operating speeds,
MLL-SGD~\cite{castiglia2021multilevel} and FedNova~\cite{Vincent_Poor2021}. 
FedNova also supports several common local optimizers, such
as SGD with proximal steps and local momentum. 
Both these works provide analysis that give insight into the benefit 
of supporting these features in federated learning algorithms.
However, these previous works in HFL cannot be applied to the VFL case.
HFL algorithms rely on distributed gradient descent methods and share model parameter updates,
while most VFL algorithms utilize distributed coordinate descent methods 
and share the output of feature extractors.
Thus, the algorithms and analyses for VFL algorithms are fundamentally different.

VFL algorithms are typically variations of 
coordinate descent methods. 
Parallel and distributed coordinate descent methods have been proposed~\cite{DBLP:conf/icml/BradleyKBG11,DBLP:conf/icml/LiuWRBS14,DBLP:journals/jmlr/RichtarikT16},
but these works depend on a shared memory structure or data sharing between parties,
which is not applicable to the VFL setting.
Several works have proposed variants of distributed coordinate descent methods
for VFL.
Many early works do not include support 
for multiple local iterations~\cite{wan2007privacy, yang2019parallel, hardy2017private}.
Without support for multiple local iterations, progress in optimization is limited by
communication time with the server, which can be costly in cases of high communication latency.
\rev{Some works  
propose synchronous VFL algorithms that support multiple local iterations~\cite{9415026, liu2019communication, castiglia2022compressed, Fu2022VFLcache, ZhangGQZWLZ22}. 
However, 
these algorithms require all parties to use the same standard SGD local optimizer.
Xie et al.~\cite{xie2022VFLADMM} propose a synchronous VFL algorithm with multiple 
local iterations using an ADMM-based optimizer. 
However, all parties use the same local optimizer in their method, 
and the fusion network is limited to a linear model, reducing the types of 
model architectures supported.
Additionally, all of these algorithms require that all parties run the same number of local iterations,
allowing stragglers to create a bottleneck in training time.}

Several works propose asynchronous VFL algorithm~\cite{FDML, chen2020vafl, ZhangGDH21, Shi2022Async}, 
but the algorithms do not support multiple local iterations.
Additionally, these algorithms only support SGD local updates. 
\rev{Gu et al.~\cite{gu2021privacy} and Zhang et al.~\cite{ZhangGDGBPH21} 
propose several asynchronous VFL algorithms that support local iterations
and common optimizers other than SGD.}
However, the schemes employed require that each party uses a linear model, 
which limits the use cases of the proposed algorithms.
\rev{Their algorithms also do not support parties using different local optimizers
in the same algorithm execution.}

\rev{In contrast with previous work, our work
jointly provides support and analyzes the effects of
party heterogeneity, time-varying speeds, and limited bandwidth in a VFL setting.}
Specifically, each party can execute a different number of 
local iterations in each round, and this number 
of local iterations can change over time. 
Further, we provide an analysis of our algorithm 
that includes the impact of this heterogeneity on 
the convergence rate and convergence error. 
These features in our algorithm and analysis allow us to model a 
more realistic VFL scenario and gives us insight into how to 
adapt party learning rates to mitigate the error introduced by party heterogeneity.

\section{System Model and Problem Formulation} \label{problem.sec}

\begin{table}
    \caption{Summary of notation used throughout the paper.}
\label{notation.table}
\vskip 0.1in
\small
\centering
\resizebox{0.49\textwidth}{!}{
    {\renewcommand{\arraystretch}{1.2}
\begin{tabular}{|l|l|}
    \hline
        \textbf{Notation} & \textbf{Definitions} \\
    \hline
    $K$ & Number of parties / Number of vertical partitions.  \\
    \hline
    $\bm x_k^i$ & Local features of data sample $i$ belonging to party $k$. \\
    \hline
    $\X_k$ & Local features belonging to party $k$ of all data samples. \\
    \hline
    $y^i$ & Label for data sample $i$. \\
    \hline
    $\y$ & Labels for all data samples. \\
    \hline
    $\bm \Theta$ & Global model parameters. \\
    \hline
    $\thetab_k$ & Party $k$'s local partition of the global model parameters. \\
    \hline
    $\bm h_k(\cdot)$ & Embedding function for party $k$. \\
    \hline
    $l_i(\cdot)$ & Loss function on data sample $i$. \\
    \hline
    $F(\cdot)$ & The objective function. \\
    \hline
    $\bm g_k(\cdot)$ & Stochastic partial derivative of the objective function with respect to $\theta_k$. \\
    \hline
    $\B$ & Mini-batch of data samples and their associated labels. \\
    \hline
    $L$,$L_k$ & Smoothness parameters for $\nabla_{\Theta}l_i(\cdot)$ 
                and $\nabla_{\theta_k}l_i(\cdot)$, respectively. \\
    \hline
    $\sigma_k$ & Variance of party $k$'s stochastic partial derivatives. \\
    \hline
    $\tau_k^r$ & Number of local iterations taken by party $k$ in round $r$. \\
    \hline
    $\eta_k^r$ & Learning rate for party $k$ in round $r$.\\
    \hline
    $\Phi_{-k}^r$ & Set of embeddings from all parties in round $r$ except party $k$. \\
    \hline
    $w_k^{r,t}$ & Weights on party $k$'s stochastic partial derivatives at local iteration $t$ and round $r$. \\
    \hline
\end{tabular}}
}
\vspace{-0.5em}
\end{table}

In this section, we present our system model and problem formulation.
We consider a system with a set of parties ${\Hm = \{1, \ldots, K\}}$.
The parties communicate via a central server, forming 
a hub-and-spoke architecture. 
The parties and the server may have different operating speeds, 
and these rates may change over time.
We will formalize party operating speeds in Section~\ref{alg.sec}.

Each party $k$ has a local dataset $\X_k \in \mathbb{R}^{N \times D_k}$.
We let the $i$-th row of $\X_k$ be denoted by $\x^i_k$. 
We assume that these local datasets are aligned, i.e., 
$\x^i_k$ and $\x^i_j$ for all parties $k \neq j$ 
are different features of the same data sample with sample ID $i$.
We let ${\X \in \mathbb{R}^{N \times D} = [\X_1, \ldots , \X_K]}$
where $D = \sum_k D_k$. 
We can see each $\X_k$ as a vertical partition of $\X$. 
Let the $i$-th data sample be the $i$-th row in $\X$, which we denote
as $\x^i$.  
Let $\y \in \mathbb{R}^{N \times 1}$ be the corresponding labels
for the data samples, and let $y^i$ be the label of the $i$-th data sample. 
We assume that the parties and server have a copy of the labels $\y$.
We discuss cases where labels are private and only present at a single party in Section~\ref{alg.sec}.

Each party $k$ holds a local model characterized by 
an \emph{embedding} function $\bm h_k(\cdot)$ and parameterized by model parameters $\thetab_k \in \mathbb{R}^{V_k}$.
An embedding function $\bm h_k$ maps the raw features $\X_k$ to a representation space, typically of lower dimension. 
For example, $\bm h_k$ can be a neural network. Each party may have a different model architecture.
We let the $k$-th \emph{embedding} of a data sample $\x^i$ be $\bm h_k(\thetab_k; \x_k^i)$, 
the output of party $k$'s feature extractor. 
If $\bm h_k$ is a neural network, then 
an embedding is the output of last layer of the network for single sample.
The server stores a \emph{server model} with parameters $\thetab_0 \in \mathbb{R}^{V_0}$. 
The server model is a function of the embeddings from each party and its output
is a predicted label $\hat{y}_i$.
We define the \emph{global model} parameters 
as $\bm\Theta = [\thetab_0 , \ldots , \thetab_K] \in \mathbb{R}^{V}$
where $V = \sum_k V_k$;
each $\thetab_k$ is a \emph{coordinate partition} of $\bm\Theta$.
The goal of the parties is to train $\bm\Theta$.
We provide an example of the VFL model structure 
in Figure~\ref{vflmodel.fig}. 
A benefit of the structure of $\bm\Theta$ is that parties can
compute partial derivatives of the loss function by exchanging
embeddings rather than exchanging their $\thetab_k$ parameters.
Since the size of the embeddings is often much smaller than 
$\thetab_k$, message sizes can be greatly reduced with this structure.
Going forward, for simplicity of notation, 
we may drop $\x^i$ from $\bm h_k(\cdot)$ when the context is clear.

The VFL objective is to minimize the following function:
\begin{align}
    F(\bm\Theta ; \X ; \y) \coloneqq 
    \frac{1}{N} \sum_{i=1}^N l_i(\thetab_0; \bm h_1(\thetab_1 ; \x_1^i) ; \ldots ; \bm h_K(\thetab_K ; \x_K^i); y^i) 
\end{align}
where $l_i(\cdot)$ is the loss function for a data sample $\x^i$
and its corresponding label $y^i$.
The loss function $l_i(\cdot)$ measures the error in predicting a label $y^i$
and can be a non-linear function,
such as cross-entropy loss, a support vector machine, or a deep neural network,
as shown in Figure~\ref{vflmodel.fig}.

Let the partial derivative associated with the coordinate partition $\thetab_k$ be:
\begin{align*}
    &\nabla_k F(\bm\Theta ; \X ; \y)  \coloneqq \\
    &~~~~~~~
    \frac{1}{N} \sum_{i=1}^N
\nabla_{\thetab_k} l_i(\thetab_0; \bm h_1(\thetab_1 ; \x_1^i) ; \ldots ; \bm h_K(\thetab_K ; \x_K^i); y^i). 
\end{align*}
Let $\B$ be a mini-batch of indices of size $B$
corresponding to a subset of rows in $\X$. 
We let $\X^{\B}$ be the rows of $\X$ that correspond to a mini-batch $\B$. 
Similarly, we let $\y^{\B}$ be the labels that correspond to $\B$.
With some abuse of notation, we define $\bm h_k(\thetab_k; \X_k^{\B})$ 
to be the set of embeddings for $\B$ for party $k$. 
We denote the stochastic partial derivative of the coordinate partition $\thetab_k$ as:
\begin{align*}
    &\bm g_k(\thetab_0 ; \bm h_1(\thetab_1; \X_1^{\B}) ; \ldots ; \bm h_K(\thetab_K; \X_K^{\B}) ; \y^{\B}) \coloneqq \\
    &~~~~~~~
    \frac{1}{B} \sum_{i \in \B} 
\nabla_{\thetab_k} l_i(\thetab_0; \bm h_1(\thetab_1 ; \x_1^i) ; \ldots ; \bm h_K(\thetab_K ; \x_K^i); y^i).
\end{align*}
With a slight abuse of notation, we let the partial derivatives 
$\bm g_k(\thetab_0 ; \bm h_1(\thetab_1; \X_1^{\B}) ; \ldots ; \bm h_K(\thetab_K; \X_K^{\B}) ; \y^{\B})$
be equivalently denoted as 
$\bm g_k(\bm\Theta ; \B)$. 
We may drop $\X$ and $\y$ from $F(\cdot)$ and $\B$ from $\bm g_k(\cdot)$ 
when the context is clear.

We make the following standard assumptions for 
$l_i(\cdot)$, $F(\cdot)$, and 
$\bm g_k(\cdot)$~\cite{bottou2018optimization, tsitsiklis1986distributed,HogWild18}:
\begin{assumption}
    \label{smooth.assum} \textbf{Smoothness}:  
    There exists positive constants ${L < \infty}$ and ${L_k < \infty}$ for $k=0,...,K$
    such that for all ${\bm\Theta_1 \in \mathbb{R}^{V}}$, $\bm\Theta_2 \in \mathbb{R}^{V}$:
        \begin{align}
            \lrVert{\nabla_{\bm\Theta} l_i(\bm\Theta_1) - \nabla_{\bm\Theta} l_i(\bm\Theta_2)} &\leq L \lrVert{\bm\Theta_1 - \bm\Theta_2} \\
            \lrVert{\nabla_{\thetab_k} l_i(\bm\Theta_1) - \nabla_{\thetab_k} l_i(\bm\Theta_2)} &\leq L_k \lrVert{\bm\Theta_1 - \bm\Theta_2}.
        \end{align}
\end{assumption}
\begin{assumption}
    \label{bias.assum} \textbf{Unbiased gradients}:  
    For a mini-batch $\B$, for $k=0,...,K$, 
        the stochastic partial derivatives are unbiased, i.e., for all $\bm\Theta \in \mathbb{R}^{V}$
        \begin{align}
            \mathbb{E}_{\B}\left[\bm g_k(\bm\Theta)\right] &= \nabla_{k} F(\bm\Theta).
        \end{align}
\end{assumption}
\begin{assumption}
    \label{var.assum} \textbf{Bounded variance}:
     There exists constants ${\sigma_k < \infty}$ for $k=0,...,K$
        such that the variances of the stochastic partial derivatives are bounded as
        \begin{align}
            \mathbb{E}_{\B}{\lrVert{\nabla_{k} F(\bm\Theta) -  \bm g_k(\bm\Theta)}^2} &\leq \sigma_k^2
        \end{align}
        for a mini-batch $\B$ and for all $\bm\Theta \in \mathbb{R}^{V}$.
\end{assumption}
Assumption~\ref{smooth.assum} bounds how fast the gradient and partial derivatives can
change. Assumption~\ref{bias.assum} requires that the stochastic 
partial derivatives computed by each party and the server
are unbiased estimates of the full-batch partial derivatives. 
Assumption~\ref{bias.assum} can be satisfied in practice by 
ensuring that sample IDs for a mini-batch are chosen at random.
Note, we make no assumption about the distribution of the full dataset $\X$.
Finally,  
Assumption~\ref{var.assum} bounds the variance between the stochastic
partial derivatives and full-batch partial derivatives by some constant.

\section{Algorithm} \label{alg.sec}

We now present Flex-VFL, our algorithm for training a global model
with distributed, vertically partitioned data in a
system with heterogeneous parties.
In each \emph{global round} of Flex-VFL, we employ a type of
parallel stochastic block coordinate descent.
Each party and the server 
updates its coordinate partition %
using its local optimizer for one or more \emph{local iterations}.
The parties complete
as many of these local iterations as possible before a specified timeout. 
This differs from synchronous VFL algorithms that wait for
all parties to complete the same number of local iterations~\cite{liu2019communication}, 
which can lead to a bottleneck when slow parties are present.
We assume that all parties participate in each global round and run at least
one local iteration before the given timeout.
Flex-VFL runs $R$ global rounds of training.

One of the challenges in training for our objective is distributing the
necessary information for parties to compute their partial derivatives.
For a given data sample $\x$ and its label $y$, 
the server must compute:
\begin{align*}
    &\bm g_0(\bm\Theta ; \x) = \nabla_{\thetab_0}l(\thetab_0 ; \bm h_1(\thetab_1; \x_1),...,\bm h_K(\thetab_K; \x_K); y) 
\end{align*}
and each party $k$ must compute: 
\begin{align*}
    &\bm g_k(\bm\Theta ; \x) = \nabla_{\thetab_k}l(\thetab_0 ; \bm h_1(\thetab_1; \x_1),...,\bm h_K(\thetab_K; \x_K); y) \\
                    &= \nabla_{\thetab_k} \bm h_k(\thetab_k)^{\top} \nabla_{\bm h_k(\thetab_k)}l(\thetab_0 ; \bm h_1(\thetab_1; \x_1),...,\bm h_K(\thetab_K; \x_K); y).
\end{align*}
We can see that for a party to calculate its model update, it
needs to first calculate 
$\nabla_{\bm h_k(\thetab_k)}l(\thetab_0 ; \bm h_1(\thetab_1; \x_1),...,\bm h_K(\thetab_K; \x_K); y)$.
Thus, to execute multiple local iterations on a data sample~$\x$ without
communication, each party and the server will need to receive a snapshot of
the embeddings $\bm h_j(\thetab_j)$ for all $j \neq k$ 
and the server model $\thetab_0$ at the start of each 
global round. 
We describe the process of exchanging this information below. 
Moving forward, the superscript $r,t$ denotes global round $r$ at local iteration $t$. 
We let the snapshot of embeddings for a mini-batch at round $r$ be denoted
$\Phi^r = \{\thetab_0^{r,0}, \bm h_1(\thetab_1^{r,0} ; \X_1^{\B}), \ldots, \bm h_K(\thetab_K^{r,0} ; \X_K^{\B})\}$.

We describe the training process of Flex-VFL in Algorithm~\ref{vafl2.alg}.
The global model is initialized to $\bm\Theta^{0,0} = [\thetab_0^{0,0}, ..., \thetab_K^{0,0}]$.
In each global round, the parties agree upon a 
mini-batch of sample IDs $\B$ on which to train.
For example, the server may assign the same random number generator seed
to all parties, ensuring each party chooses the same sample IDs at the start of each round.
Each party then determines its set of local features $\X_k^{\B}$ 
that it will use to compute embeddings.
Each party, excluding the server, computes its embeddings for the mini-batch, $\bm h_k(\thetab_k; \X^{\B}_k)$. 
These embeddings are sent to the server, and the server distributes
its current model $\thetab_0$ and $\Phi^r$ to all parties.

\begin{algorithm}[t]
    \begin{algorithmic}[1]
        \State {\textbf{Initialize:}} $\thetab_k^{0,0}$ for all parties $k$ and server model $\thetab_0^{0,0}$
        \For {$r \leftarrow 0, \ldots, R-1$}
            \State Select a mini-batch of sample IDs $\B^r \in \{\X, \y\}$
            \For {$k \leftarrow 1, \ldots, K$ in parallel}
                \State Sample features $\X_k^{\B^r}$ corresponding to IDs in $\B^r$
                \State Send $\bm h_k(\thetab_k^{r,0} ; \X^{\B^r}_k)$ to server %
            \EndFor
            \State $\Phi^r \leftarrow \{\thetab_0^{r,0}, \bm h_1(\thetab_1^{r,0} ; \X_1^{\B^r}), \ldots, \bm h_K(\thetab_K^{r,0} ; \X_K^{\B^r})\}$ 
            \State Server sends $\Phi^r$ to all parties
            \For {$k \leftarrow 0, \ldots, K$ in parallel}
                \LineComment \rev{\textit{Parties and server run until local training timeout}}
                \For {$t \leftarrow 0, \ldots, \tau_k^r-1$}
                    \State $\thetab_k^{r,t+1} \leftarrow \thetab_k^{r,t} - \eta_k^{r}                         
                    \textbf{D}_k\left(\Phi_{-k}^r ; \thetab_k^{r,0}, \ldots, \thetab_k^{r,t} ; \y^{\B}\right)$ %
                \EndFor
                \State $\thetab_k^{r+1,0} \leftarrow \thetab_k^{r,\tau_k^r}$
            \EndFor
        \EndFor
    \end{algorithmic}
    \caption{Flexible Vertical Federated Learning}
    \label{vafl2.alg}
\end{algorithm}

\rev{Next, each party and the server start the \emph{local training period}.
Each party and the server run as many local gradient descent steps as it can within 
a timeout that is pre-specified before training begins.
We let $\tau_k^r$ be the number of local gradient descent steps that party $k$
completes in round $r$, with $1 \leq \tau_k^r < \infty$.
This number of local iterations $\tau_k^r$ depends on 
party $k$'s operating speed in global round $r$ and the local training timeout.}
Slower parties have smaller $\tau_k^r$ values, 
while faster parties have larger $\tau_k^r$ values.
Since computation loads may change over time, a party's operating speed,
and thus its $\tau_k^r$, may change between each global round.
Note that during local iterations, each party only updates its own embedding $\bm h_k(\thetab_k^{r,t})$,
and uses stale versions of $\bm h_j(\thetab_j^{r,0})$ for all $j \neq k$. 
Each party also reuses the same mini-batch for $\tau_k^r$ iterations, saving on communication of a new set
of embeddings at each local iteration.
We show in Section~\ref{adapt.sec} that the model converges, even though
this stale information is used.

A party $k$'s updates during each local iteration are defined as follows:
\begin{align}
\thetab_k^{r,t+1} = \thetab_k^{r,t} - \eta_k^{r}                         
\textbf{D}_k\left(\Phi_{-k}^r ; \thetab_k^{r,0}, \ldots, \thetab_k^{r,t} ; \y^{\B}\right)  
    \label{update.eq}
\end{align}
where $\Phi_{-k}^r$ is the set of all embeddings except those from party $k$, 
and $\textbf{D}_k(\cdot)$ is the local optimizer update rule based on the stochastic
partial derivatives from the start of the global round to the current local iteration.
The cumulative update to each local model over 
a set of local iterations is as follows:
\begin{align}
    \thetab_k^{r,\tau_k^r} &= 
    \thetab_k^{r,0} - \eta_k^{r}                         
    \sum_{t=0}^{\tau_k^r-1} \textbf{D}_k\left(\Phi_{-k}^r ; \thetab_k^{r,0}, \ldots, \thetab_k^{r,t} ; \y^{\B}\right) 
    \label{update2.eq}
\end{align}
We assume that the cumulative update over all local iterations 
can be rewritten in the following form:
\begin{align}
    &\sum_{t=0}^{\tau_k^r-1} \textbf{D}_k\left(\Phi_{-k}^r ; \thetab_k^{r,0}, \ldots, \thetab_k^{r,t} ; \y^{\B}\right) 
    \coloneqq 
    \nonumber \\ &~~~~~~~~~~~~~~~~~~~~~~~~~~~~
    \sum_{t=0}^{\tau_k^r-1} w_k^{r,t} \bm g_k(\Phi_{-k}^r ; \bm h_k(\thetab_k^{r,t}) ; \y^{\B})  
    \label{updateD.eq}
\end{align}
where $w_k^{r,t} \geq 1$ are weights applied to each gradient update.
Note that we do not assume that each index of these sums in (\ref{updateD.eq})
are equivalent.
By rewriting party update rules in this way, we can analyze a variety
of common local optimizers.
Below, we present some examples of common local optimizers whose updates can be
written in the form of (\ref{updateD.eq}).
\begin{itemize}[leftmargin=*]
    \item \textbf{Classical SGD:} 
        In classical SGD, our update rule is: 
        \begin{align*}
            \textbf{D}_k\left(\Phi_{-k}^r ; \thetab_k^{r,0}, \ldots, \thetab_k^{r,t} ; \y^{\B}\right)                          
            &\coloneqq \bm g_k(\Phi_{-k}^r ; \bm h_k(\thetab_k^{r,t}) ; \y^{\B}). 
        \end{align*}
        By setting $w_k^{r,t}=1$ for all parties, rounds, and local iterations in
        (\ref{updateD.eq}), our update rule becomes classical SGD.
    \item \textbf{SGD with Momentum:} 
        SGD with local momentum is where a party resets its momentum buffer to zero 
        at the start of each global round.
        The update rule for SGD with local momentum is defined as follows:
        \begin{align*}
            \textbf{D}_k\left(\Phi_{-k}^r ; \thetab_k^{r,0}, \ldots, \thetab_k^{r,t} ; \y^{\B}\right) 
            &\coloneqq u_k^{r,t}  
        \end{align*}
        where
        \begin{align*}
            u_k^{r,0} &= \bm g_k(\Phi_{-k}^r ; \bm h_k(\thetab_k^{r,0})) \\ 
            u_k^{r,t} &= \rho u_k^{r,t-1} + \bm g_k(\Phi_{-k}^r ; \bm h_k(\thetab_k^{r,t})) 
        \end{align*}
        where $\rho$ is a tunable parameter.
        The updates $u_k^{r,t}$ at each local iteration $t$ can be defined as follows:
        \begin{align*}
            u_k^{r,t} 
            &= \rho u_k^{r,t-1} + \bm g_k(\Phi_{-k}^r ; \bm h_k(\thetab_k^{r,t-1})) \\
            &= \rho^2 u_k^{r,t-2} + \rho \bm g_k(\Phi_{-k}^r ; \bm h_k(\thetab_k^{r,t-2})) 
            \nonumber \\ &~~~~~~~~~~~~~~~~~~~~~~~~~
            + \bm g_k(\Phi_{-k}^r ; \bm h_k(\thetab_k^{r,t-1})) \\
            &= \sum_{s=0}^{t-1} \rho^{t-1-s} \bm g_k(\Phi_{-k}^r ; \bm h_k(\thetab_k^{r,t-1})).
        \end{align*}
        Plugging this into (\ref{update.eq}), we have:
        \begin{align*}
            \thetab_k^{r,t} &= \thetab_k^{r,t-1} - \eta_k^r \sum_{s=0}^{t-1} \rho^{t-1-s} \bm g_k(\Phi_{-k}^r ; \bm h_k(\thetab_k^{r,s}))\\
            &= \thetab_k^{r,t-2} - \eta_k^r \sum_{s=0}^{t-2} \rho^{t-2-s} \bm g_k(\Phi_{-k}^r ; \bm h_k(\thetab_k^{r,s})) 
            \nonumber \\ &~~~~~~~~~~~~~~~~
            - \eta_k^r \sum_{s=0}^{t-1} \rho^{t-1-s} \bm g_k(\Phi_{-k}^r ; \bm h_k(\thetab_k^{r,s})).
        \end{align*}
        Applying this recursion to our update rule in (\ref{update2.eq}), we have:
        \begin{align*}
            \thetab_k^{r,\tau_k^r} = \thetab_k^{r,0} - \eta_k^r \sum_{t=0}^{\tau_k^r-1} \sum_{s=0}^{t} \rho^{t-s} \bm g_k(\Phi_{-k}^r ; \bm h_k(\thetab_k^{r,s})).
        \end{align*}
        Thus, in order to represent local momentum, the weights in (\ref{updateD.eq}) can be set as follows:
        \begin{align*}
        w_k^{r,t} = 1+\rho+\rho^2+\cdots+\rho^{\tau_k^r-1-t} = \frac{1-\rho^{\tau_k^r-t}}{1-\rho}.
        \end{align*}
    \item \textbf{Proximal updates:}  
        As first proposed in FedProx~\cite{li2018federated}, 
        and first applied to a VFL algorithm by Liu et al.~\cite{liu2019communication},
        one can apply a proximal term by defining a party's update rule as follows:
        \begin{align*}
            &\textbf{D}_k\left(\Phi_{-k}^r ; \thetab_k^{r,0}, \ldots, \thetab_k^{r,t} ; \y^{\B}\right) 
            \coloneqq \\ &~~~~~~~~~~~~~~~~~~~~~~~~~~
            \bm g_k(\Phi_{-k}^r ; \bm h_k(\thetab_k^{r,t})) 
            + \mu \left(\thetab_k^{r,t} - \thetab_k^{r,0}\right)
        \end{align*}
        where $\mu$ is a tunable parameter.
        Plugging this into (\ref{update.eq}) we have:
        \begin{align*}
            \thetab_k^{r,t+1}
            = \thetab_k^{r,t} - \eta_k^r \left(
            \bm g_k(\Phi_{-k}^r ; \bm h_k(\thetab_k^{r,t})) 
            + \mu \left(\thetab_k^{r,t} - \thetab_k^{r,0}\right) \right)
        \end{align*}
        
        Subtracting $\thetab_k^{r,0}$ from both sides we have: 
        \begin{align*}
            \thetab_k^{r,t+1} - \thetab_k^{r,0}
            &= (1-\eta_k^r \mu)\left(\thetab_k^{r,t} - \thetab_k^{r,0}\right) 
            \\ &~~~~~~~~~~~~~~~
            - \eta_k^r \bm g_k(\Phi_{-k}^r ; \bm h_k(\thetab_k^{r,t})).
        \end{align*}
        Repeatedly applying the recursion on $\thetab_k^{r,t}$, we have:
        \begin{align*}
            \thetab_k^{r,t+1}
            &= \thetab_k^{r,0}
            - \eta_k^r \sum_{t=0}^{\tau_k^r-1} (1-\eta_k^r \mu)^{\tau_k^r-1-t}
            \bm g_k(\Phi_{-k}^r ; \bm h_k(\thetab_k^{r,t})) .
        \end{align*}
        Thus, the weights in (\ref{updateD.eq}) can be set as follows to
        represent proximal steps in VFL:
        \begin{align*}
            w_k^{r,t} = (1-\eta_k^r \mu)^{\tau_k^{r,t}-t+1}.
        \end{align*}
\end{itemize}
This method of generalizing to several local optimizers was 
first shown in the context of HFL by Wang et al.~\cite{Vincent_Poor2021}. 
However it has yet to be analyzed in the context of VFL, 
which provides its own unique challenges.
We discuss this more in Section~\ref{adapt.sec}.

\textit{Communication cost:}
The size of messages in Flex-VFL is of note. 
For each party to compute its partial derivative,
every party must exchange its embeddings for the current mini-batch, 
and the server must send its model $\thetab_0$ to the parties.
Let the size of the $k$-th embedding for a single data sample be $O_k$.
The total amount of data sent per global round is
then $K(|\thetab_0|+B\sum_{k=1}^KO_k)$. 

\textit{Privacy:}
HFL algorithms typically share model updates or gradient information in messages.
Gradients can potentially leak raw data information, as shown in previous 
work~\cite{DBLP:journals/tifs/PhongAHWM18, DBLP:conf/nips/GeipingBD020}.
However, in Flex-VFL, messages only contain embeddings and each party
can only calculate the partial derivatives associated with the
server model and its local model. 
Thus, gradient attacks proposed for HFL cannot be performed on Flex-VFL.
Embeddings may be vulnerable to model inversion attacks~\cite{MahendranV15},
though these attacks can be mitigated by applying
homomorphic encryption~\cite{SecureBoost, hardy2017private} 
or secure multi-party computation~\cite{gu2021privacy} to Flex-VFL.
As mentioned in Section~\ref{problem.sec}, 
we assume that all parties have access to the labels.
There are many practical scenarios where data samples are private between
the parties, but the labels are not, such as predicting credit score.
However, if labels are private and only present at a single party, 
Flex-VFL can be augmented using the method proposed by Liu et al.~\cite{liu2019communication},
allowing gradient calculation without the need for sharing labels.
The analysis in Section~\ref{adapt.sec} still holds in this case.

\section{Analysis} \label{adapt.sec}
In this section, we provide convergence analysis of Flex-VFL.
To avoid cumbersome notation going forward, we define 
$\bm g_k^{r,t} \coloneqq \bm g_k(\Phi_{-k}^r; \bm h_k(\thetab_k^{r,t}; \X_k^{\B}) ; \y^{\B})$
and drop $\Phi_{-k}^r$ from $\textbf{D}_k(\cdot)$ when the context is clear.

We start by defining a recurrence relation for updates to the global model.
Let $G^r$ be the stack of gradient updates in a global round $r$:
\begin{align}
    \G^r 
    &= \left[\sum_{t_0=0}^{\tau_0^r-1} \textbf{D}_0(\thetab_0^{r,0},\ldots,\thetab_0^{r,t_0}), \ldots, \sum_{t_K=0}^{\tau_K^r-1}\textbf{D}_K(\thetab_K^{r,0},\ldots \thetab_K^{r,t_K}) \right] \nonumber \\
    &= \left[\sum_{t_0=0}^{\tau_0^r-1} w_0^{r,t_0} \bm g_0^{r,t_0}, \ldots, \sum_{t_K=0}^{\tau_K^r-1} w_K^{r,t_K} \bm g_K^{r,t_K}\right].
\end{align}
We can define our updates to the global model during a global round 
with the following recurrence relation:
\begin{align}
    \bm\Theta^{r+1,0} = \bm\Theta^{r,0} - \eta_k^r \G^r.
    \label{recursion.eq}
\end{align}
With the help of (\ref{recursion.eq}), we can model
Flex-VFL as a gradient coordinate descent algorithm 
and analyze the algorithm convergence in this vein.

\rev{We note that each party reuses stale embeddings of the same mini-batch from other parties
for multiple local iterations in Algorithm~\ref{vafl2.alg}:
each party $k$ takes $\tau_k^r$ descent steps using mini-batch $\B^r$ 
at a global round $r$.
This indicates that the stochastic gradients are not unbiased during local iterations $t > 0$.
However, using conditional expectation, we can apply Assumption~\ref{bias.assum} 
to the gradient calculated at local iteration $t=0$.}
If we take expectation over $\B^r$, conditioned
on the previous models $\bm\Theta^{\tau,0}$ up to round $r$, we obtain
\begin{align}
    &\mathbb{E}_{\B^r}[\bm g_k^{r,0} ~|~ \{\bm\Theta^{\tau,0}\}_{\tau=0}^{r}] = \nabla_m F( \Phi_{-k}^r ; \bm h_k(\thetab_k^{r,t}) ).
    \label{bias_imp2.assum}
\end{align}
With the help of (\ref{bias_imp2.assum}), we can prove convergence by
bounding the difference between the gradient at 
the start of each global round and those calculated during local iterations.

In particular, we prove the following lemma:
\begin{lemma} \label{diff.lemma}
If $\eta_k^r \leq \frac{1}{2 \tau_k^r L_k \max_{0 \leq t \leq \tau_k^r-1}w_k^{r,t}}$,
    then under Assumptions~\ref{smooth.assum}--\ref{var.assum} 
    the weighted conditional expected squared norm difference of 
    gradients $\bm g_k^{r,t}$ and $\bm g_k^{r,0}$ 
    for a set of $\tau_k^r$ local iterations is bounded as follows:
\begin{align}
    & \sum_{t=0}^{\tau_k^r-1} w_k^{r,t} \Ebatch{\lrVert{\bm g_k^{r,t} - \bm g_k^{r,0}}^2}
    \nonumber \\ &
    \leq 8(\tau_k^r)^3(\eta_k^r)^2 L_k^2 \max_{0 \leq t \leq \tau_k^r-1} (w_k^{r,t})^3 \left(\lrVert{\nabla_k F(\bm\Theta^{r,0})}^2 + \sigma_k^2\right)
\end{align}
and
\begin{align}
    &\sum_{t=0}^{\tau_k^r-1} (w_k^{r,t})^2 \Ebatch{\lrVert{\bm g_k^{r,t} - \bm g_k^{r,0}}^2}
    \nonumber \\ &
    \leq 8(\tau_k^r)^3(\eta_k^r)^2 L_k^2 \max_{0 \leq t \leq \tau_k^r-1} (w_k^{r,t})^4 \left(\lrVert{\nabla_k F(\bm\Theta^{r,0})}^2 + \sigma_k^2\right)
\end{align}
where $\mathbb{E}^r$ is the expectation taken on the mini-batch
$\B^r$ conditioned on $\{\bm\Theta^{\tau,0}\}_{\tau=0}^r$. 
\end{lemma}
The proof of Lemma~\ref{diff.lemma} can be found in Appendix~\ref{lemma.sec}.
\rev{Lemma~\ref{diff.lemma} bounds the error incurred at each party in the 
stochastic partial derivatives during a set of local iterations
as a result of using stale embeddings from other parties, 
as well as from using the same mini-batch for all local iterations in a single round.
The lemma also captures the effect of different local optimizers at each party on
the partial derivatives.
Using Lemma~\ref{diff.lemma}, we can now analyze how this error accumulates 
over all iterations.}
Applying our smoothness assumption along with (\ref{recursion.eq}) and 
Lemma~\ref{diff.lemma}, we can prove that Flex-VFL converges.%

We present our convergence result in the following theorem.
\begin{theorem} \label{main.thm}
    Under Assumptions~\ref{smooth.assum}--\ref{var.assum}, if $\eta_k^r$ satisfies:
\begin{align}
    \eta_k^r \leq \frac{1}{16 \tau_k^r \max\{L, L_k\} \underset{0 \leq t \leq \tau_k^r-1}{\max} w_k^{r,t}}
    \label{constraint.eq}
\end{align}
then the weighted average squared gradient norm over all parties and 
$R$ rounds of Algorithm~\ref{vafl2.alg} is bounded by:
\begin{align}
    & 
    \frac{1}{S}\sum_{r=0}^{R-1} \sum_{k=0}^K \eta_k^r W_k^r 
    \Etot{\lrVert{\nabla_k F(\bm\Theta^{r,0})}^2}
    \leq 
    \frac{4}{S}(F(\bm\Theta^{0,0}) - \Finf)  
    \nonumber \\ &~~~~~~~~~~~~~~
    + \frac{4L}{S} \sum_{r=0}^{R-1} 
    \sum_{k=0}^K (\eta_k^r)^2 W_k^r \max_{0 \leq t \leq \tau_k^r-1} w_k^{r,t} \tau_k^r \sigma_k^2  
    \label{main.eq}
\end{align}
where $\Finf$ is a lower bound on $F(\cdot)$,
$S = \sum_{r=0}^{R-1} \sum_{k=0}^K \eta_k^r W_k^r$,
and $W_k^r = \sum_{t=0}^{\tau_k^r-1} w_k^{r,t}$.
\end{theorem}
We provide the full proof of Theorem~\ref{main.thm} in Appendix~\ref{proof.sec}.

The left-hand-side of (\ref{main.eq}) is a weighted average
of all the partial derivatives' norms during the training.
As this term approaches zero, Flex-VFL approaches a fixed point.
The first term in the right-hand-side of (\ref{main.eq}) is similar to that of 
distributed gradient descent~\cite{bottou2018optimization},
and is affected by the difference in the initial and final models of the algorithm.
The first term is the convergence rate term:
as the number of global rounds $R$ approaches $\infty$, this first term goes to zero, 
while the second term, the additive convergence error, remains.
The second term is the error associated with the variance in
taking stochastic gradients steps, as well as the error incurred by
running multiple local iterations. 

If we let $\eta_k^r=\frac{1}{\sqrt{R\max_{k,r}\tau_k^r}}$ in (\ref{main.eq}),
then we can see the convergence rate of Flex-VFL is $O(\frac{1}{\sqrt{R\max_{k,r}\tau_k^r}})$,
the same as other VFL algorithms~\cite{
liu2019communication,chen2020vafl, FDML}.
This indicates that we can achieve a fast convergence speed despite the error introduced by
heterogeneous party speeds and local optimizers.

\textbf{Effect of heterogeneous speeds:}
We observe that $S$ appears in the denominator of the first term in~(\ref{main.eq});
a larger value of $S$ improves the convergence rate of Flex-VFL.
This quantity $S$ depends on the party learning rates $\eta_k^r$ and 
the constraint (\ref{constraint.eq}) requires that $\eta_k^r$ 
be inversely proportional to $\tau_k^r$, 
the number of local iterations party $k$ takes in round $r$. 
In addition, $S$ increases with the sum of weights $w_k^{r,t}$ over $\tau_k^r$ local iterations.
Consider the case where each party employs classical 
SGD, implying $w_k^{r,t} = 1$ for all parties over all iterations.
In this case, $S = \sum_{r=0}^{R-1} \sum_{k=0}^K \eta_k^r \tau_k^r$.
If we let $\eta_k^r = \frac{1}{\tau_k^r}$, then $S = R(K+1)$.
If we instead consider a fixed learning rate $\eta_k^r = \eta$ 
across all parties and iterations,
then to satisfy (\ref{constraint.eq}), $\eta = \frac{1}{\max_{k,r}\tau_k^r}$.
In this case, $S = \sum_{r=0}^{R-1} \sum_{k=0}^K \frac{\tau_k^r}{\max_{k,r}\tau_k^r} \leq R(K+1)$.
Thus, choosing the learning rates $\eta_k^r$ according to 
each party's number of local iterations in a round $\tau_k^r$ 
can reduce the error in the first term of (\ref{main.eq})
and improve the convergence rate.
However, $\tau_k^r$ may not always be known in advance.
We discuss these findings further in Section~\ref{adapt2.sec},
where we introduce Adaptive Flex-VFL.

\textbf{Effect of heterogeneous optimizers:}
From Theorem~\ref{main.thm}, we can see 
that the first term in (\ref{main.eq}) decreases with larger 
local optimizer weights, while the second term increases with larger weights.
Note that proper tuning of $\eta_k^r$ for each party $k$ 
can help offset the error introduced
by $\max_{0 \leq t \leq \tau_k^r-1} w^t$ in the second term.
We also note that the constraint (\ref{constraint.eq}) 
requires $\eta_k^r$ to be inversely proportional $\max_{0 \leq t \leq \tau_k^r-1} w^t$.
Thus there is tension between choosing larger optimizer weights and
choosing a larger step size. 
In cases where $\eta_k^r$ remains constant while still satisfying (\ref{constraint.eq}),
Theorem~\ref{main.thm} indicates that weights larger than $1$, such as when
using momentum and proximal steps, can improve the convergence rate
by decreasing the first term in (\ref{main.eq}).
However, it can also negatively affect the error as $R \rightarrow \infty$
by increasing the second term in (\ref{main.eq}). 
If all else is constant, 
a large $W_k^r$ is beneficial when $\sigma_k$ is small.
In other words, in cases where stochastic variance is small,
our analysis shows a potential benefit for using local optimizers
other than classical SGD.

We now introduce some corollaries to 
study the convergence rate of Flex-VFL.
We first consider the case where each party has the same stochastic variance,
runs the same number of local iterations, and uses SGD locally.
\begin{corollary} \label{main1.cor}
    Suppose $\sigma_k = \sigma$, $\tau_k^r = \tau$, and $\eta_k^r = \eta$ 
    for all parties $k$ and rounds $r$.
    Let $w_k^{r,t}=1$ for all rounds $r$, local iterations $t$, and parties $k$.
    Under Assumptions~\ref{smooth.assum}-\ref{var.assum}, if $\eta$ satisfies:
\begin{align}
    \eta \leq \frac{1}{16 \tau \max\{L, L_k\}}
\end{align}
then the average squared gradient norm over all parties and 
$R$ rounds of Algorithm~\ref{vafl2.alg} is bounded by:
\begin{align}
    \frac{1}{R} \sum_{r=0}^{R-1} \Etot{\lrVert{\nabla F(\bm\Theta^{r,0})}^2} 
    &\leq 
    \frac{4(F(\bm\Theta^{0,0}) - \Finf)}{R \eta \tau(K+1)}  
    + 4L \eta \tau \sigma^2 
    \label{simple.eq}
\end{align}
\end{corollary}

If we let $\eta=\frac{1}{\sqrt{\tau R}}$ in Corollary~\ref{main1.cor}, 
then we can see our
convergence rate is $O(\frac{1}{\sqrt{\tau R}})$,
which is the same as distributed SGD algorithms~\cite{
DBLP:journals/siamjo/GhadimiL13a, bottou2018optimization, ZhouC18Kstep}.

We consider a decaying learning rate in the following corollary.
\begin{corollary} \label{main2.cor}
    Suppose 
    $\eta_k^r \leq \frac{1}{16 \tau_k^r \max\{L, L_k\} \max_{0 \leq t \leq \tau_k^r-1} w_k^{r,t}}$,
    and suppose that
    $\sum_{r=0}^{\infty} \eta_k^r = \infty$ and
    $\sum_{r=0}^{\infty} (\eta_k^r)^2 < \infty$ for all parties $k$.
    Then under Assumptions~\ref{smooth.assum}-\ref{var.assum}, 
    the left-hand-side of (\ref{main.eq}) goes to zero as $R$ approaches $\infty$.
\end{corollary}
Corollary~\ref{main2.cor} states that given a sequence of learning rates that
are not summable, but square summable, then Algorithm~\ref{vafl2.alg}
    achieves convergence to a fixed point.
    One possible choice for learning rates to satisfy these conditions 
    is by diminishing $\eta_k^r$ at a rate of $O\left( \frac{1}{r} \right)$ 
    where $r$ is the current global round. This is a standard step size
    requirement of SGD algorithms for non-convex objectives~\cite{bottou2018optimization}.

\section{Adaptive Extension} \label{adapt2.sec}

In this section, we present an adaptive extension to Flex-VFL,
which we call Adaptive Flex-VFL.
One can think of Adaptive Flex-VFL as a meta-optimization algorithm
where the server keeps track of party operating speeds and local optimizer parameters
during each global round in order to choose the best learning rates for convergence speed.

\rev{Let $\bar{w}_k^r \coloneqq \underset{0 \leq t \leq \tau_k^r-1}{\max} w_k^{r,t}$.}
According to Theorem~\ref{main.thm}, the learning rate $\eta_k^r$
at a given global round $r$ for party $k$ must be inversely 
proportional to the number of local iterations $\tau_k^r$
and the largest weight applied during local iterations
$\bar{w}_k^r$. 
We also know that the first term of (\ref{main.eq}) 
is the main contributor to the convergence rate in Flex-VFL, 
while the second term is the convergence floor,
mostly affected by the stochastic variance.
If the bound (\ref{constraint.eq}) on the learning rates $\eta_k^r$ holds,
the second term's effect is minimal.
Therefore, based on our analysis, a natural improvement to Flex-VFL
is to maximize $\eta_k^r$ subject to (\ref{constraint.eq}) in each global round,
tailoring each party's learning rate to their specific operating rate and local
optimizer parameters.

\begin{algorithm}[t]
    \begin{algorithmic}[1]
        \State {\textbf{Initialize:}} $\thetab_k^{1,0}$ for all parties $k$ and server model $\thetab_0^{1,0}$ 
        \State \rev{{\textbf{Initialize:}} $A_k$ and $\eta_k^1 = \frac{A_k}{\tau_k^0 \bar{w}_k^0}$ for all parties}
        \For {$r \leftarrow 1, \ldots, R$}
            \State Sample a mini-batch $\B^r \in \{\X, \y\}$
            \For {$k \leftarrow 1, \ldots, K$ in parallel}
                \State Sample features $\X_k^{\B^r}$ corresponding to IDs in $\B^r$
                \State Send $\bm h_k(\thetab_k^{r,0} ; \X^{\B^r}_k)$ to server %
            \EndFor
            \State $\Phi^r \leftarrow \{\thetab_0^{r,0}, \bm h_1(\thetab_1^{r,0}), \ldots, \bm h_K(\thetab_K^{r,0})\}$ 
            \State Server sends $\Phi^r$ to all parties
            \For {$k \leftarrow 0, \ldots, K$ in parallel}
                \For {$t \leftarrow 0, \ldots, \tau_k^r-1$}
                    \State $\thetab_k^{r,t+1} \leftarrow \thetab_k^{r,t} - \eta_k^{r}                         
                    \textbf{D}_k(\Phi_{-k}^r ; \bm h_k(\thetab_k^{r,t}) ; \y^{\B})$ %
                \EndFor
                \State \rev{Send $\tau_k^r$ and $\bar{w}_k^r$ to server}
            \EndFor
            \State \rev{Server sets $\eta_k^{r+1} \leftarrow \frac{A_k}{\tau_k^r \bar{w}_k^r}$ for all parties}
        \EndFor
    \end{algorithmic}
    \caption{Adaptive Flexible Vertical Federated Learning}
    \label{adapt.alg}
\end{algorithm}

Adaptive Flex-VFL is outlined in Algorithm~\ref{adapt.alg}.
For the first global round, $\tau_k^0$ for each party is estimated, 
either with prior knowledge of the party operating speeds, 
or by running a dummy global round.
Each party can communicate $\bar{w}_k^0$ to the server before the start of training.
All $\eta_k^1$ are initialized to some $\frac{A_k}{\tau_k^0 \bar{w}_k^0}$, where
$A_k$ is chosen small enough to satisfy (\ref{constraint.eq}).
At the end of each global round, the server gathers
information about how many local iterations each party took, $\tau_k^r$,
and their maximum weight applied to gradients during local iterations, $\bar{w}_k^r$.
We let $\eta_k^{r+1} = \frac{A_k}{\tau_k^r \bar{w}_k^r}$.
If we assume that for each party $k$, 
$\tau_k^r$ and $w_k^{r,t}$ do not change too rapidly across global rounds,
the server can accurately estimate the appropriate learning rate 
to assign to each party that will maximize the convergence rate.

\rev{We now define more formal conditions under which
Adaptive Flex-VFL is guaranteed to converge according to Theorem~\ref{main.thm}.
Suppose the maximum rate of change of $\tau_k^r \bar{w}_k^r$
from a round $r$ to $r+1$ is $\alpha$:
\begin{align}
    \frac{\tau^{r} \bar{w}_k^r}{\tau^{r+1} \bar{w}_k^{r+1}} &\leq \alpha.
\end{align}
Note that $\eta_k^{r+1} = \frac{A_k}{\tau^{r} \bar{w}_k^{r}}$.
In order to satisfy constraint \eqref{constraint.eq}, we need $\eta_k^{r+1} \leq \frac{1}{16 \max\{L, L_k\} \tau^{r+1} \bar{w}_k^{r+1}}$. 
Therefore, we need $A_k$ to be chosen such that:
\begin{align}
\frac{A_k}{\tau^{r} \bar{w}_k^{r}}
    &\leq \frac{1}{16 \max\{L, L_k\} \tau^{r+1} \bar{w}_k^{r+1}} \\
    A_k
    &\leq \alpha \left( \frac{1}{16 \max\{L, L_k\}}\right).
    \label{ck.eq}
\end{align}
Thus, in order for Adaptive Flex-VFL to satisfy \eqref{constraint.eq} in all rounds of training,
$A_k$ must satisfy~\eqref{ck.eq}.
We note that in practice, as we show in Section~\ref{exp.sec}, 
Adaptive Flex-VFL can show a clear improvement over Flex-VFL with a fixed learning rate.
}

\section{Experiments} \label{exp.sec}

Next, we present experiments to compare Flex-VFL
with synchronous and asynchronous VFL algorithms, and
to observe the effect of the adaptive extension 
to Flex-VFL in practice.\footnote{
Code for experiments available at \url{https://github.com/rpi-nsl/flex-vfl}}

\subsection{Datasets and Experimental Setup}

We utilize three datasets for our experiments: 
the MOSEI dataset~\cite{MOSEI2018},
the ImageNet dataset~\cite{imagenet},
and the ModelNet40 dataset~\cite{wu20153d}.
For each dataset and VFL algorithm, we performed a grid search to choose 
the best learning rate and regularization parameters (where applicable);
we trained each algorithm with different hyperparameters
for $100$ epochs and chose the hyperparameters with the lowest training loss.

\textbf{MOSEI:}
CMU-MOSEI is a multimodal dataset for sentiment analysis. 
The dataset consists of $23$,$453$ sentences from YouTube videos
giving opinions on various topics. The dataset includes video, audio,
and text data, and each sentence is labeled with sentiment values, 
scoring the positivity or negativity of the sentence.
For our experimental setup, we consider a case with three parties,
where each party stores one type of data.
The parties with video and audio train local LSTMs, and the party with
text trains a BERT model. The server model consists of a 
three-layer fully-connected neural network. 
We use an L1 loss function with L2 regularization. 
The parties train using SGD with a batch size of $50$.
For the video and audio parties the learning rate was selected from
$\{0.01, 0.005, 0.001, 0.0005\}$. For the text party the learning rate
was selected from $\{\num{0.00005}, \num{0.00001}, \num{0.000005}\}$ %
The regularization coefficient was selected from $\{0,10^{-3},10^{-4},10^{-5}\}$. 

\textbf{ImageNet100:}
The ImageNet dataset consists of images from several classes of objects.
In our experiments, we randomly choose $100$ classes from the ImageNet dataset (ImageNet100),
consisting of ${\sim}130$,$000$ images. 
We consider a case with $4$ parties, each storing a quadrant of each image.
Each party trains ResNet50 locally and the server trains a single fully connected layer.
We use a cross-entropy loss function with L2 regularization. 
The parties train using a batch size of $256$.
Each party trains using SGD with local momentum with $\rho=0.9$.
The initial learning rate for each party was selected from
$\{0.3, 0.08, 0.03, 0.008, 0.003\}$ and the regularization 
coefficient was selected from $\{0,10^{-3},10^{-4},10^{-5}\}$. 
The learning rate decays by a factor of $10$ every ${\sim}75$,$000$ local iterations.

\textbf{ModelNet40:}
The ModelNet40 dataset are images of CAD models with $40$ classes of objects, 
each with $12$ different camera views.
In our experiments, we consider a setup with $12$ parties, 
each with a single view of each CAD model.
Each party trains ResNet50 locally and the server trains a single fully connected layer.
We use a cross-entropy loss function for training. 
The parties train using a batch size of $64$.
Each party trains using SGD with local momentum with $\rho=0.9$.
The learning rate for each party was selected from
$\{1\times10^{-3}, 5\times10^{-4}, 1\times10^{-4}, 5\times10^{-5}\}$. 

\textbf{Baselines:}
\rev{We compare Flex-VFL with synchronous and asynchronous VFL methods. 
We limit our comparisons to those that support arbitrary party feature
extractors and arbitrary server fusion networks, as well as 
multiple local iterations~\cite{castiglia2022compressed, liu2019communication, chen2020vafl}.}
\begin{itemize}[leftmargin=*]
    \item \textbf{Sync-VFL} Synchronous VFL is a special case of Flex-VFL where all parties run the same number of local iterations, regardless of party operating speeds.
        When parties all use standard SGD and the server model has no trainable parameters, 
        Sync-VFL is equivalent to the VFL algorithm proposed by Liu et al~\cite{liu2019communication}.
        For our experiments, we consider two cases of Sync-VFL: Sync-Min-VFL and Sync-Max-VFL. 
    \begin{itemize}
        \item \textbf{Sync-Min-VFL:} Flex-VFL and Sync-VFL use the same choice of local training timeout,
    meaning each party in Sync-VFL runs $\tau^r = \min_{k} \tau_k^r$ 
    descent steps before synchronizing. 
        \item \textbf{Sync-Max-VFL:} Sync-VFL waits for all parties to run 
    $\tau^r = \max_{k} \tau_k^r$ descent steps before synchronizing.
    This means that Sync-Max-VFL ensures that all parties run the same number 
    of iterations as the fastest party in Flex-VFL, which extends the duration of a global round
            to accommodate the slowest party.
    \end{itemize}
\item \textbf{VAFL:} In VAFL~\cite{chen2020vafl}, each party calculates 
its embeddings for a randomly selected mini-batch.
The party then immediately exchanges information with the server
to update both the server model and party's model parameters.
Parties may take different lengths of time to execute their individual training steps, 
and so the model updates are asynchronous.
\item \textbf{P-BCD:} We also include a baseline for Flex-VFL where we set 
$\tau_k^r=1$ for all $k$ and $r$.
This is equivalent to parallel block coordinate descent (P-BCD)~\cite{SmithFMTJJ17}.
\end{itemize}

\textbf{Time units:}
In each of our experiments, we measure time-to-target in terms of simulated time units. 
We define the communication time with the server,
the computation time for each party,
and the local training timeout to be used by Flex-VFL and Sync-VFL 
in terms of these time units. 
The time units taken to complete a local iteration and 
the timeout chosen inform how many local iterations each 
party will perform in Flex-VFL and Sync-VFL.
For example, suppose two parties take $5$ and $10$ time units to complete a local iteration, 
respectively, and the timeout is set to $20$.
In Sync-VFL, both parties will complete two local iterations.
In Flex-VFL, the first party will complete four local iterations while the second party completes two.

\subsection{Heterogeneous Optimizers}

\begin{table}[t]
    \caption{Time to reach target mean-squared error (MAE) on the MOSEI
    using different optimizer combinations. 
The value shown is the mean of $5$ runs, $\pm$ the standard deviation.
    }
\label{hetero.table}
\vskip 0.1in
\small
\centering
\resizebox{0.49\textwidth}{!}{
\begin{tabular}{llc}
    \toprule 
    \multirow{2}{*}{\begin{tabular}{@{}c@{}} \textbf{Video/Audio} \\ \textbf{Optimizers} \end{tabular} } 
        & \multirow{2}{*}{\begin{tabular}{@{}c@{}} \textbf{Text} \\ \textbf{Optimizer} \end{tabular} } &
             \multirow{2}{*}{\textbf{Time units ($\times 10^3$) to reach target}} \\
             && \\
    \midrule
    SGD & SGD                       & 64.55 $\pm$ 11.74   \\
    Local Momentum & SGD            & 35.21 $\pm$ 11.74   \\
    SGD & Local Momentum            & \textbf{29.34 $\pm$ 0.00}  \\
    Local Momentum & Local Momentum & 41.08 $\pm$ 14.37   \\
                                                
    \bottomrule
\end{tabular}}
\vspace{-0.5em}
\end{table}

We first study the effect of training with heterogeneous optimizers.
We run Flex-VFL and train on the MOSEI dataset with parties either using
standard SGD or SGD with local momentum.
\rev{We consider a case where all parties and the server have the 
same operating rate.
We let the computation time for each party be $1$ time unit 
and set the timeout for each global round to be $20$ time units.
Thus, each party runs $20$ local iterations in each global round.
We let the communication latency be $10$ time units.}
We measure the average time taken to achieve a target mean-square error (MAE)
over $5$ runs.
The results are given in Table~\ref{hetero.table}. 
The best optimizer combination over the five runs is using
standard SGD at the video and audio parties that train LSTMs, while
the text party uses SGD with local momentum to train the BERT model.
Thus we can see a benefit from choosing different optimizers at each
party depending on the local model architecture.
We use this combination of optimizers for the rest of the experiments with the MOSEI dataset.

\subsection{Fixed Operating Speeds}

We next study the setting where party operating speeds 
are heterogeneous but remain fixed throughout training.
For each dataset we define the speed of a party by how many time units 
it takes for it to complete a local iteration, as well as a local training timeout
to be used by Flex-VFL and Sync-VFL.

\rev{For MOSEI, we set the timeout to $20$ time units, and
set the operating rates such that 
the parties storing video, audio, and text take
$5$, $10$, and $15$ local iterations before the timeout, respectively, 
while the server completes $20$ local iterations before the timeout.
For ImageNet100, we set the timeout to $10$ time units, and set 
the operating rates such that
the four parties take $2$, $4$, $6$, and $8$ local iterations before the timeout, 
respectively, while the server completes $10$ local iterations.
For ModelNet40, we set the timeout to $20$ time units,  
set the operating rates such that server take $20$ local iterations before the timeout, and
let groups of $3$ parties each take $5$, $10$, $15$, and $20$ local iterations, respectively.
These operating rates are chosen such that there are an equal number of stragglers, 
medium speed parties, and fast parties in the system.}

We simulate three communication network settings, representing cases of
different ratios of computation time versus communication time.
We denote the round-trip message latency with the server as $t_{\mathrm{comm}}$ and
we let the time for the fastest party to complete a local iteration be $1$ time unit.
In the first setting, we assume communication latency with the server is very low,
equal to computation time of a single local iteration, and let $t_{\mathrm{comm}} = 1$ unit. 
In the second setting, we consider a case where communication time 
starts to outweigh computation time of a single local iteration, 
and set $t_{\mathrm{comm}} = 10$ units. In this case, communication with the server takes 
$10$ local iterations at the fastest party.
For the final setting, we consider the case where there is very high communication latency, 
setting $t_{\mathrm{comm}} = 50$ units.
Such high communication latency can occur when parties are globally distributed.

\begin{table}
    \caption{Time units taken to reach target test accuracy
    under different communication times.
For the MOSEI dataset, the target is reaching $0.65$ MAE.
For ImageNet100, target is reaching $60\%$ top-$5$ accuracy.
For ModelNet40, the target is reaching $70\%$ top-$5$ accuracy.
The value shown is the mean of $5$ runs, $\pm$ the standard deviation.
A ``--" indicates that the target was not reached during training.}
\label{acc.table}
\vskip 0.1in
\small
\centering
\resizebox{0.49\textwidth}{!}{
\begin{tabular}{llccc}
    \toprule 
    \multirow{3}{*}{\begin{tabular}{@{}c@{}} \textbf{Comm.} \\ \textbf{Time}\end{tabular}} & \multirow{3}{*}{\textbf{Algorithm}} & \multicolumn{3}{c}{\textbf{Time units ($\times 10^3$) to reach target}} \\
                      \cmidrule(rl){3-5}
    && \subhead{MOSEI} & \subhead{ImageNet100} & \subhead{ModelNet40}\\
    && \subhead{Target=0.65 MAE} & \subhead{Target=60\% Top-$5$ Acc.} & \subhead{Target=70\% Top-$5$ Acc.}\\
    \midrule
    \multirow{4}{*}{1 unit} &P-BCD       & \textbf{5.22 $\pm$ 0.65} & \textbf{73.06 $\pm$ 2.38}   & 44.51 $\pm$ 0.90      \\
                            &VAFL        & 7.63 $\pm$ 1.12          & 89.59 $\pm$ 8.67            & 52.27 $\pm$ 2.17       \\
                            &Sync-Min-VFL& 13.69 $\pm$ 0.00         & 299.48 $\pm$ 19.17          & 32.34 $\pm$ 2.05     \\
                            &Sync-Max-VFL& 26.41 $\pm$ 0.00         & 520.05 $\pm$ 63.06          & 84.82 $\pm$ 4.99     \\
                            &Flex-VFL    & 6.85 $\pm$ 0.00          & 113.26 $\pm$ 31.07          & \textbf{25.23 $\pm$ 2.42}     \\
    \midrule
  \multirow{4}{*}{10 units} &P-BCD       & 14.60 $\pm$ 1.83         & \textbf{182.66 $\pm$ 5.94}   & 124.62 $\pm$ 2.51    \\
                            &VAFL        & 29.60 $\pm$ 3.18         & --                          & 161.02 $\pm$ 3.93     \\ 
                            &Sync-Min-VFL& 19.56 $\pm$ 0.00         & 544.50 $\pm$ 34.86          & 46.20 $\pm$ 2.92     \\
                            &Sync-Max-VFL& 29.34 $\pm$ 0.00         & 611.82 $\pm$ 74.19          & 94.25 $\pm$ 5.54     \\
                            &Flex-VFL    & \textbf{9.78 $\pm$ 0.00} & 205.92 $\pm$ 56.49          & \textbf{36.04 $\pm$ 3.46}     \\
    \midrule
  \multirow{4}{*}{50 units} &P-BCD       & 56.33 $\pm$ 7.04         & 669.74 $\pm$ 21.78           & 480.66 $\pm$ 9.70    \\
                            &VAFL        & 96.09 $\pm$ 6.41         & --                           & --                  \\
                            &Sync-Min-VFL& 45.64 $\pm$ 0.00         & 1633.50 $\pm$ 104.58        & 107.80 $\pm$ 6.82     \\
                            &Sync-Max-VFL& 42.38 $\pm$ 0.00         & 1019.70 $\pm$ 123.65        & 136.14 $\pm$ 8.01     \\
                            &Flex-VFL    & \textbf{22.82 $\pm$ 0.00}& \textbf{617.76 $\pm$ 169.47}& \textbf{84.08 $\pm$ 8.07}     \\
    \bottomrule
\end{tabular}}
\vspace{-0.5em}
\end{table}

In Table~\ref{acc.table} we show the time units it takes to reach a target test accuracy.
For the MOSEI dataset, we let the target be $0.65$ Mean Absolute Error (MAE),
a common measure of performance for the dataset.
We can see that for the MOSEI dataset, P-BCD and VAFL  
perform best when communication latency is low. 
However, as communication latency increases, Flex-VFL is able to 
reach the target MAE twice as fast as Sync-Min-VFL and P-BCD,
and four times as fast as VAFL.
For the ImageNet100 dataset, the target is set to $60\%$ top-$5$ accuracy.
In this case, P-BCD performs well when communication latency is low.
As the communication latency increases however, Flex-VFL overtakes P-BCD,
benefiting from local iterations saving on overall communication.
Finally, for the ModelNet40 dataset, the target is set to $70\%$.
Here, Flex-VFL always performs better than the other algorithms,
regardless of communication latency. 
In our experiments with ModelNet40, there are a larger number of 
parties than in the other datasets. 
This causes the heterogeneity of party operating speeds to 
have a greater effect on time-to-target accuracy.
With the chosen distribution of party operating speeds, Flex-VFL
reaches a target accuracy up to four times as fast than other VFL algorithms.
In the next subsection, we consider how different operating speed 
distributions can affect performance of each VFL algorithm.

\subsection{Uniform Versus Average Operating Rates} 
For our next experiment, we again consider fixed operating speeds, 
and now focus on the effect of operating rate distribution.
Specifically, we compare an evenly distributed case of party operating speeds to a typical case,
and show how Flex-VFL, Sync-VFL, VAFL, and P-BCD perform.
For these experiments, we use the ModelNet40 dataset,
and consider two distributions of party operating speeds. 
For Flex-VFL and Sync-VFL, we use a local training timeout of $20$ time units.
The first distribution is the same as the previous experiment:
\rev{the timeout is set to $20$ time units,
and the operating rates are set such that
the server takes $20$ local iterations per round, and
groups of $3$ parties each take $5$, $10$, $15$, and $20$ local iterations, respectively.}
For determining a realistic typical case of party operating speeds, we use 
Google's Cluster Data~\cite{clusterdata:Reiss2011}, a dataset of workload traces
running on Google compute cells. 
We choose $13$ random traces from the dataset, and take the average CPU usage
for each to determine the operating speed of parties for these experiments.
For a party $k$, we let the time to complete a local iteration be 
$(1-c_k)^{-1}$ for all $r$ rounds, 
where $c_k$ is the $k$-th machine's average CPU utilization. 
\rev{Since the timeout is $20$ time units, $\tau_k^r = 20\cdot(1-c_k)$ for all $r$ rounds.}

\begin{figure}[t]
    \begin{subfigure}{0.23\textwidth}
        \centering
        \includegraphics[width=\textwidth]{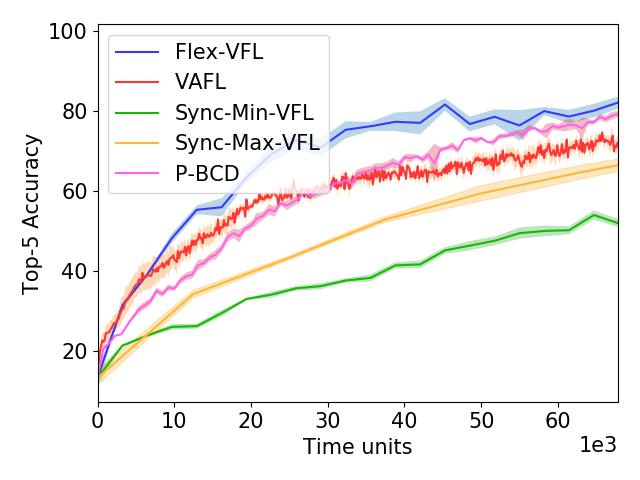}
        \caption{$t_{\mathrm{comm}}=1$ with evenly distributed operating speeds.}
        \label{vafl1acc.fig}
    \end{subfigure}
    \hfill
    \begin{subfigure}{0.23\textwidth}
        \centering
        \includegraphics[width=\textwidth]{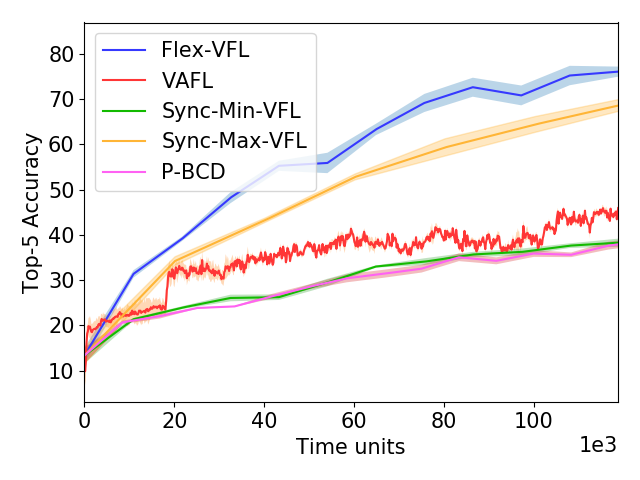}
        \caption{$t_{\mathrm{comm}}=50$ with evenly distributed operating speeds.}
        \label{vafl50acc.fig}
    \end{subfigure}
    \begin{subfigure}{0.23\textwidth}
        \centering
        \includegraphics[width=\textwidth]{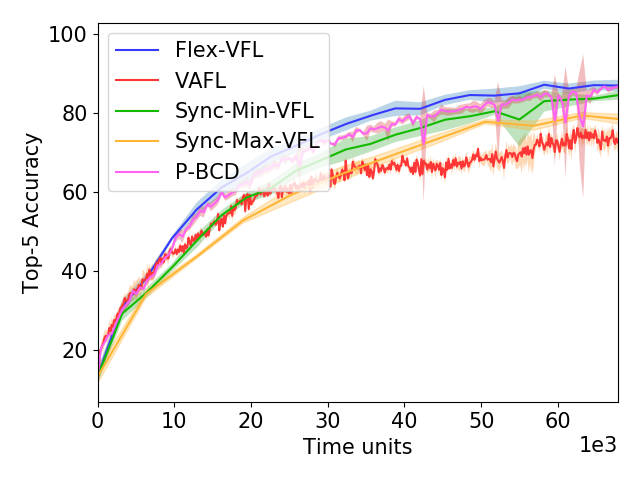}
        \caption{$t_{\mathrm{comm}}=1$ with Google Cluster operating rate distribution.}
        \label{vafl1acc2.fig}
    \end{subfigure}
    \hfill
    \begin{subfigure}{0.23\textwidth}
        \centering
        \includegraphics[width=\textwidth]{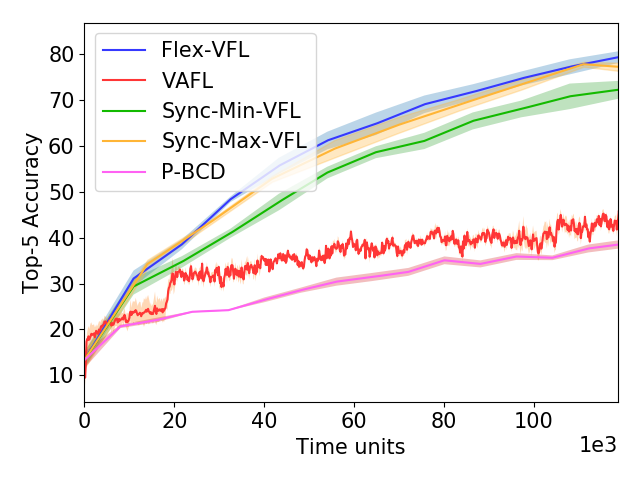}
        \caption{$t_{\mathrm{comm}}=50$ with Google Cluster operating rate distribution.}
        \label{vafl50acc2.fig}
    \end{subfigure}
    \caption{Top-$5$ test accuracy plotted against time units for ModelNet40 dataset
    with uniform and average distribution of party operating speeds.
    The solid lines are the mean of $5$ runs, while the shaded region represents
    the standard deviation.}
    \label{vafl.fig}
\end{figure}

In Figure~\ref{vafl.fig} we present the results of training with the different operating rate distributions.
We plot the top-$5$ test accuracy against time units. 
In Figures~\ref{vafl1acc.fig} and \ref{vafl50acc.fig}, we see the results when
party operating speeds are evenly distributed. 
In this case, we can see that when communication time is low, Flex-VFL and VAFL
perform well, as they are robust to the stragglers in the system.
P-BCD also performs well here. We believe this is because P-BCD 
does not incur error introduced by local iterations.
When communication latency is high though, we can see that Flex-VFL continues to perform
well while VAFL and P-BCD start to perform worse. 
In Figures~\ref{vafl1acc2.fig} and \ref{vafl50acc2.fig}, we see the results for
operating speeds using the Google Cluster data. 
In this case, we can see that Flex-VFL performs the same or better than Sync-Min-VFL and P-BCD
when communication latency is low. When communication latency is high, 
Flex-VFL performs the same or better than Sync-Max-VFL. 
Flex-VFL in these cases is always the best choice of algorithm regardless
of party operating rate distributions or communication time.

\rev{
\subsection{Effect of Local Training Timeout}
In this section, we explore the effect of different timeouts with fixed operating rates
on the time-to-target accuracy of Flex-VFL.
For these experiments, we use the ModelNet40 dataset
and use the even distribution of operating rates from previous experiments.
We let the communication time be $10$ time units.
We also include results from Sync-Max-VFL as a baseline.
Recall that Sync-Max-VFL ensures all parties to run the same number of local iterations
as the fastest party does in Flex-VFL. 
For example, if the fastest party in Flex-VFL runs $20$ local iterations,
then all parties in Sync-Max-VFL run $20$ local iterations.

\begin{table}
    \caption{\rev{Time units taken to reach $70\%$ top-$5$ test accuracy 
    on the ModelNet40 dataset.
    On the left, we include the time-to-target accuracy for Flex-VFL for timeouts of $20$ to $40$ units.
    On the right, we include the time-target accuracy of Sync-VFL for $20$ to $40$ local iterations.
    For example, in row $1$, the fastest party in Flex-VFL runs $20$ local iterations, while
    all parties in Sync-VFL run $20$ local iterations.
    Communication time is $10$ units.
    The value shown is the mean of $5$ runs, $\pm$ the standard deviation.}}
\label{timeout.table}
\vskip 0.1in
\small
\centering
    \rev{
\resizebox{0.24\textwidth}{!}{
\begin{tabular}{ccc}
    \toprule 
    \multirow{3}{*}{\begin{tabular}{@{}c@{}} \textbf{Local Training} \\ \textbf{Period Timeout}\end{tabular}} & \textbf{Flex-VFL} \\
                      \cmidrule(rl){2-2}
    &  Time units ($\times 10^3$)\\
    & to reach target\\
    \midrule
    $20$ units & \textbf{36.04 $\pm$ 3.46} \\
    $25$ units & \textbf{30.49 $\pm$ 3.70} \\
    $30$ units & \textbf{24.02 $\pm$ 1.85} \\
    $35$ units & \textbf{22.18 $\pm$ 1.85} \\
    $40$ units & \textbf{33.26 $\pm$ 1.85} \\
    \bottomrule
\end{tabular}}
\resizebox{0.24\textwidth}{!}{
\begin{tabular}{ccc}
    \toprule 
    \multirow{3}{*}{\begin{tabular}{@{}c@{}} \textbf{Local Iterations} \\ \textbf{Per Round}\end{tabular}} & \textbf{Sync-VFL} \\
                      \cmidrule(rl){2-2}
    &  Time units ($\times 10^3$)\\
    & to reach target\\
    \midrule
    $20$ iterations & 94.25 $\pm$ 5.54 \\
    $25$ iterations & 72.07 $\pm$ 5.54 \\
    $30$ iterations & 60.98 $\pm$ 6.79 \\
    $35$ iterations & 58.21 $\pm$ 5.54 \\
    $40$ iterations & 85.93 $\pm$ 5.54 \\
    \bottomrule
\end{tabular}}}
\end{table}

Table~\ref{timeout.table} shows the time taken for Sync-VFL and Flex-VFL to 
achieve $70\%$ top-$5$ accuracy on the ModelNet40 dataset for different timeouts.
We see that for Flex-VFL, increasing the timeout improves the time-to-target, up until
a timeout of $40$ time units. 
We see this same trend with Sync-VFL when changing the number of local iterations.
This is in line with previous works that explore
the effect of local iterations on convergence rate~\cite{liu2019communication, lin2018don, pmlr-v54-mcmahan17a}. 
Increasing the number of local iterations improves time-to-target up to a
certain point, where the error incurred by local iterations outweighs the benefits to
convergence rate.
We can also see in Table~\ref{timeout.table} that Flex-VFL outperforms Sync-VFL
regardless of the timeout.
}

\subsection{Variable Operating Speeds}

\begin{figure}[t]
    \centering
    \includegraphics[width=0.35\textwidth]{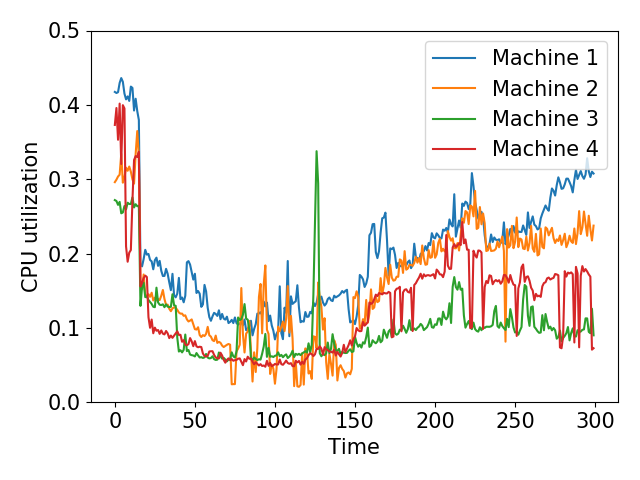}
    \caption{CPU utilization over time of $4$ machines from
        the Google Cluster workload dataset.}
    \label{cpus.fig}
\end{figure}

\begin{figure}[t]
    \begin{subfigure}{0.23\textwidth}
        \centering
        \includegraphics[width=\textwidth]{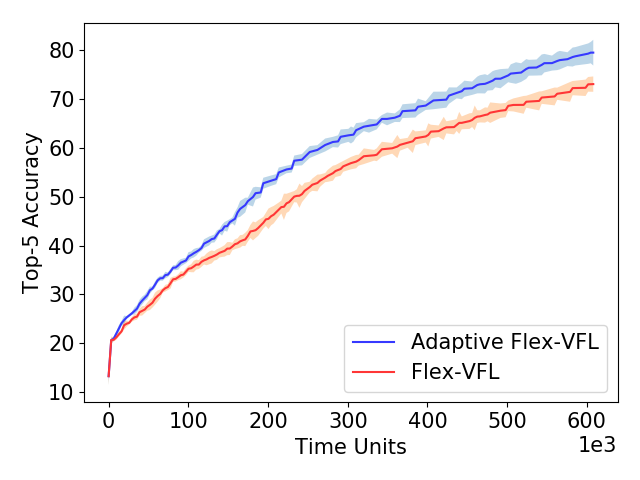}
        \caption{ModelNet40 dataset}
        \label{adapt.fig}
    \end{subfigure}
    \begin{subfigure}{0.23\textwidth}
        \centering
        \includegraphics[width=\textwidth]{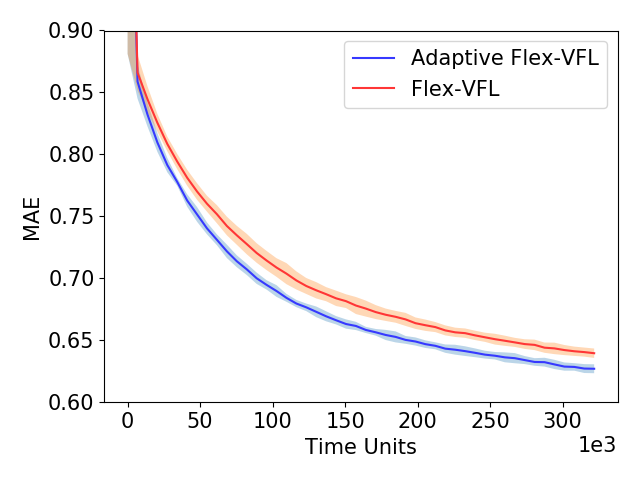}
        \caption{MOSEI dataset}
        \label{adapt2.fig}
    \end{subfigure}
    \caption{Top-$5$ test accuracy and test mean absolute error (MAE) 
    of adaptive and static methods of Flex-VFL training on the ModelNet40 and MOSEI datasets
    respectively.}
\end{figure}

We next investigate the setting where party operating speeds change during training.
To model realistic changes in these rates, we again use 
Google's Cluster Data~\cite{clusterdata:Reiss2011}.
In Figure~\ref{cpus.fig}, we plot the CPU utilization of $4$ randomly chosen machine traces
over a short time interval.
For a party $k$, we let the time to complete a local iteration be 
$(1-c_k^r)^{-1}$ where $c_k^r$ is the $k$-th machine's CPU utilization at global round $r$. 
\rev{We set $\tau_k^r = t_o \cdot(1-c_k^r)$ for round $r$, where $t_o$ is the local training timeout.}

We compare Flex-VFL and Adaptive Flex-VFL, and train on the ModelNet40 and MOSEI datasets.
For ModelNet40, we use a timeout of $10$ time units and set $A_k = 0.0001$, 
and for MOSEI, we use a timeout of $20$ time units and set $A_k = \num{0.00005}$ 
for the text party and $A_k = 0.001$ for the other parties and server.
In both cases, we let the communication time be $1$ time unit.
For static Flex-VFL, the server chooses $\eta_k = \eta_k^r$ such that
(\ref{constraint.eq}) is not violated at any point during training. 
This is determined using the largest value $\tau_k^r$ over all rounds $r$, which
is $10$ for ModelNet40 and $20$ for MOSEI, 
corresponding to the case when the CPU utilization is $c_k^r=0$.
Adaptive Flex-VFL, on the other hand, chooses $\eta_k^r$ in each global round
based on the previous rounds values of $\tau_k^r$ and $\max_{0 \leq t \leq \tau_k^r-1} w_k^{r,t}$, as described
in Algorithm~\ref{adapt.alg}.

In Figures~\ref{adapt.fig} and \ref{adapt2.fig}, 
we show the top-$5$ test accuracy of training on the ModelNet40
and MOSEI datasets, respectively.
We can see that
Adaptive Flex-VFL reaches a much higher test accuracy 
and lower MAE in a shorter period of training than Flex-VFL. 
In ModelNet40, Adaptive Flex-VFL reaches a top-$5$ accuracy of
$70\%$ about $130$ time units faster than Flex-VFL.
In MOSEI, Adaptive Flex-VFL achieves an MAE of $0.65$ about $70$ time
units faster than Flex-VFL.
Allowing the learning rates to increase when
CPU utilization is lower can greatly improve the convergence
rate of Flex-VFL.

\section{Conclusion} \label{conclusion.sec} 
We proposed Flex-VFL, a vertical federated learning algorithm 
that learns on distributed, vertically-partitioned data in a
system with heterogeneous parties. 
We analyzed Flex-VFL and showed the benefit of optimizers that
utilize momentum or proximal steps in VFL settings.
We also showed that convergence requires that each party's
learning rate is tailored to its operating speed and local optimizer.
Based on this observation, we proposed Adaptive Flex-VFL,
which optimizes party learning rates at each global round based on 
party operating speeds and optimizer parameters.
In our experiments, 
we demonstrated that Flex-VFL can outperform both synchronous and asynchronous VFL algorithms,
reaching a target accuracy up to $4\times$ faster than other VFL algorithms. 
Our experiments also indicate that Flex-VFL is often the best overall choice
of VFL algorithms when it's necessary to be flexible with high communication latency 
and different party operating rate distribution.
We also provided experimental results comparing Adaptive Flex-VFL and Flex-VFL
using real-world party operating speeds.
We found that Adaptive Flex-VFL can improve time-to-target accuracy by $30\%$ over Flex-VFL.
In future work, we will explore the impact of 
partial participation of VFL parties on algorithm performance.

\section*{Acknowledgment}
This material is based upon work supported in part by the National Science Foundation under Grant CNS-1553340, and 
 by the Rensselaer-IBM AI Research Collaboration (\url{http://airc.rpi.edu}), part of the IBM AI Horizons Network.

\appendix
In this section, we provide our full proof of Theorem~\ref{main.thm}.
We start by introducing some additional notation and providing the proof of Lemma~\ref{diff.lemma}.

\subsection{Additional Notation} \label{additional.sec}
We define
\begin{align}
    \gamma_{k,j}^{r,t} = 
    \begin{cases}
        \thetab_j^{r,t} & k = j \\
        \thetab_j^{r,0} & \text{otherwise}
    \end{cases}
\end{align}
to represent party $k$'s view of party $j$'s model in round $r$ and iteration $t$.
We define the column vector $\Gamma_k^{r,t} = [\gamma_{k,0}^{r,t};...;\gamma_{k,K}^{r,t}]$
to be party $k$'s view of the system model in round $r$ and iteration $t$.

\subsection{Proof of Lemma~\ref{diff.lemma}} \label{lemma.sec}

We now prove Lemma~\ref{diff.lemma}, stated in Section~\ref{adapt.sec}.
\begin{proof}
We start by bounding the expected squared norm difference between 
gradients at the start of the global round $r$ and local iteration $t$:
\begin{align}
    &\Ebatch{\lrVert{\bm g_k^{r,t} - \bm g_k^{r,0}}^2}  \nonumber \\
    &= \Ebatch{\lrVert{\bm g_k^{r,t} - \bm g_k^{r,t-1} + \bm g_k^{r,t-1} - \bm g_k^{r,0}}^2} \\
    &\leq (1+n)\Ebatch{\lrVert{\bm g_k^{r,t} - \bm g_k^{r,t-1}}^2} 
    \nonumber \\ &~~~~~~~~~~~~~~~~~~~~~
    + \left(1+\frac{1}{n}\right) \Ebatch{\lrVert{\bm g_k^{r,t-1} - \bm g_k^{r,0}}^2}
    \label{nplus.eq}
\end{align}
where (\ref{nplus.eq}) follows from the fact that 
$(X+Y)^2 \leq (1+n)X^2+(1+\frac{1}{n})Y^2$ for some positive $n$.

Applying Assumption~\ref{smooth.assum} to the first term in (\ref{nplus.eq}) we have:
\begin{align}
    &\Ebatch{\lrVert{\bm g_k^{r,t} - \bm g_k^{r,0}}^2}  \nonumber \\
    &\leq (1+n)L_k^2 \Ebatch{\lrVert{\Gamma_k^{r,t} - \Gamma_k^{r,t-1}}^2} 
    \nonumber \\ &~~~~~~~~~~~~~~~~~~~~~
    + \left(1+\frac{1}{n}\right) \Ebatch{\lrVert{\bm g_k^{r,t-1} - \bm g_k^{r,0}}^2} \\
    &= (1+n)(\eta_k^r)^2 L_k^2 (w_k^{r,t-1})^2 \Ebatch{\lrVert{\bm g_k^{r,t-1}}^2} 
    \nonumber \\ &~~~~~~~~~~~~~~~~~~~~~
    + \left(1+\frac{1}{n}\right) \Ebatch{\lrVert{\bm g_k^{r,t-1} - \bm g_k^{r,0}}^2} 
    \label{recursion1.eq}
\end{align}
where (\ref{recursion1.eq}) follows from the update rule 
$\thetab_k^{r,t} = \thetab_k^{r,t-1} - \eta_k^r w_k^{r,t-1} \bm g_k^{r,t-1}$.

We now add and subtract $\bm g_k^{r,0}$ to the first term in (\ref{recursion1.eq}):
\begin{align}
    &\Ebatch{\lrVert{\bm g_k^{r,t} - \bm g_k^{r,0}}^2} \nonumber \\
    &\leq (1+n)(\eta_k^r)^2 L_k^2(w_k^{r,t-1})^2 
    \Ebatch{\lrVert{\bm g_k^{r,t-1} - \bm g_k^{r,0} + \bm g_k^{r,0}}^2} 
    \nonumber \\ &~~~~~~~~~~~~~~~~~~~~~
    + \left(1+\frac{1}{n}\right) \Ebatch{\lrVert{\bm g_k^{r,t-1} - \bm g_k^{r,0}}^2} \\
    &\leq 2(1+n)(\eta_k^r)^2 L_k^2(w_k^{r,t-1})^2 
    \Ebatch{\lrVert{\bm g_k^{r,t-1} - \bm g_k^{r,0}}^2} 
    \nonumber \\ &~~~
    +2(1+n)(\eta_k^r)^2 L_k^2(w_k^{r,t-1})^2 
    \Ebatch{\lrVert{\bm g_k^{r,0}}^2} 
    \nonumber \\ &~~~
    + \left(1+\frac{1}{n}\right) \Ebatch{\lrVert{\bm g_k^{r,t-1} - \bm g_k^{r,0}}^2}. 
    \label{need_n.eq}
\end{align}

    If we let $n = \tau_k^r$, we can bound (\ref{need_n.eq}) further:
\begin{align}
    &\Ebatch{\lrVert{\bm g_k^{r,t} - \bm g_k^{r,0}}^2} \nonumber \\
    &\leq 2(1+\tau_k^r)(\eta_k^r)^2 L_k^2(w_k^{r,t-1})^2 
    \Ebatch{\lrVert{\bm g_k^{r,t-1} - \bm g_k^{r,0}}^2} 
    \nonumber \\ &~~~
    +2(1+\tau_k^r)(\eta_k^r)^2 L_k^2(w_k^{r,t-1})^2 
    \Ebatch{\lrVert{\bm g_k^{r,0}}^2} 
    \nonumber \\ &~~~
    + \left(1+\frac{1}{\tau_k^r}\right) \Ebatch{\lrVert{\bm g_k^{r,t-1} - \bm g_k^{r,0}}^2}. 
    \label{definedn.eq}
\end{align}

Let $\eta_k^r \leq \frac{1}{2\tau_k^r L_k \max_{0 \leq t \leq \tau_k^r-1}w_k^{r,t}}$. 
We bound (\ref{definedn.eq}) as follows:
\begin{align}
    &\Ebatch{\lrVert{\bm g_k^{r,t} - \bm g_k^{r,0}}^2} \nonumber \\
    &\leq \left(1+\frac{2}{\tau_k^r}\right)
    \Ebatch{\lrVert{\bm g_k^{r,t-1} - \bm g_k^{r,0}}^2} 
    \nonumber \\ &~~~
    +2(1+\tau_k^r)(\eta_k^r)^2 L_k^2 \max_{0 \leq t \leq \tau_k^r-1} (w_k^{r,t})^2 
    \Ebatch{\lrVert{\bm g_k^{r,0}}^2}. 
\end{align}

We define the following notation for simplicity: 
\begin{align}
    A^{r,t} &\coloneqq \Ebatch{\lrVert{\bm g_k^{r,t} - \bm g_k^{r,0}}^2} \\
    B &\coloneqq 2(1+\tau_k^r)(\eta_k^r)^2 L_k^2 \max_{0 \leq t \leq \tau_k^r-1} (w_k^{r,t})^2 \Ebatch{\lrVert{\bm g_k^{r,0}}^2} \\
    C &\coloneqq \left(1+\frac{2}{\tau_k^r}\right).
\end{align}

Note that we have shown that $A^{r,t} \leq CA^{r,t-1} + B$.
Utilizing this bound, we can also show that:
\begin{align}
    A^{r,1} \leq~& CA^{r,0} + B \\
    A^{r,2} \leq~& C^2A^{r,0} + CB + B \\
    A^{r,3} \leq~& C^3A^{r,0} + C^2B + CB + B \\
            &\vdots \nonumber \\
    A^{r,t} \leq~& C^tA^{r,0} + B\sum_{\tau_1=0}^{t-1} C^{\tau_1}.
    \label{recurse2.eq}
\end{align}

Note that $A^{r,0}=\Ebatch{\lrVert{\bm g_k^{r,0} - \bm g_k^{r,0}}^2}=0$.
It is left to bound the second term in (\ref{recurse2.eq}) 
over the set of local iterations.
\begin{align}
    \sum_{t=0}^{\tau_k^r-1} &w_k^{r,t}B\sum_{\tau_1=0}^{t-1} C^{\tau_1}  \nonumber \\
    &= B\sum_{t=0}^{\tau_k^r-1} w_k^{r,t}
    \left(\frac{C^{t}-1}{C-1}\right) \\
    &= \frac{B}{C-1}\sum_{t=0}^{\tau_k^r-1} w_k^{r,t}
    \left(C^{t}-1\right) \\
    &= \frac{B}{C-1}\max_{0 \leq t \leq \tau_k^r-1} w_k^{r,t}
    \left(\frac{C^{\tau_k^r}-1}{C-1}-\tau_k^r\right)  \\
    &= \frac{B}{C-1}\max_{0 \leq t \leq \tau_k^r-1} w_k^{r,t}
    \left(\frac{\left(1+\frac{2}{\tau_k^r}\right)^{\tau_k^r}-1}{\frac{2}{\tau_k^r}}-\tau_k^r\right)  \\
    &\leq \frac{(\tau_k^r)^2B}{2}\max_{0 \leq t \leq \tau_k^r-1} w_k^{r,t}
    \left(\frac{e^2-1}{2}-1\right)  \\
    &\leq 2(\tau_k^r)^2B\max_{0 \leq t \leq \tau_k^r-1} w_k^{r,t}.
    \label{need_B.eq}
\end{align}

Plugging the definition of $B$ into (\ref{need_B.eq}):
\begin{align}
    &\sum_{t=0}^{\tau_k^r-1} w_k^{r,t}B\sum_{\tau_1=0}^{t-1} C^{\tau_1} \nonumber \\
    &\leq 4(\tau_k^r)^2(1+\tau_k^r)(\eta_k^r)^2 L_k^2\max_{0 \leq t \leq \tau_k^r-1} (w_k^{r,t})^3 \Ebatch{\lrVert{\bm g_k^{r,0}}^2} \\
    &\leq 8(\tau_k^r)^3(\eta_k^r)^2 L_k^2 \max_{0 \leq t \leq \tau_k^r-1} (w_k^{r,t})^3 \Ebatch{\lrVert{\bm g_k^{r,0}}^2}.
    \label{need_as3.eq}
\end{align}

Applying Assumption~\ref{var.assum} to (\ref{need_as3.eq}):
\begin{align}
    &\sum_{t=0}^{\tau_k^r-1} w_k^{r,t} \Ebatch{\lrVert{\bm g_k^{r,t} - \bm g_k^{r,0}}^2} \nonumber \\
    &\leq 8(\tau_k^r)^3(\eta_k^r)^2 L_k^2 \max_{0 \leq t \leq \tau_k^r-1} (w_k^{r,t})^3 \left(\lrVert{\nabla_k F(\bm\Theta^{r,0})}^2 + \sigma_k^2\right).
\end{align}

Similarly:
\begin{align}
    &\sum_{t=0}^{\tau_k^r-1} (w_k^{r,t})^2 \Ebatch{\lrVert{\bm g_k^{r,t} - \bm g_k^{r,0}}^2} \nonumber \\
    &\leq 8(\tau_k^r)^3(\eta_k^r)^2 L_k^2 \max_{0 \leq t \leq \tau_k^r-1} (w_k^{r,t})^4 \left(\lrVert{\nabla_k F(\bm\Theta^{r,0})}^2 + \sigma_k^2\right).
\end{align}
This completes the proof of Lemma~\ref{diff.lemma}.

\end{proof}

\subsection{Proof of Theorem~\ref{main.thm}} \label{proof.sec}
Applying our smoothness assumption, given in Assumption~\ref{smooth.assum}:
\begin{align}
    &F(\bm\Theta^{r+1,0}) - F(\bm\Theta^{r,0}) \nonumber \\
    &\leq \left\langle \nabla F(\bm\Theta^{r,0}), \bm\Theta^{r+1,0} - \bm\Theta^{r,0} \right\rangle 
    + \frac{L}{2} \lrVert{\bm\Theta^{r+1,0} - \bm\Theta^{r,0}}^2 \\
    &\leq -\sum_{k=0}^K \eta_k^r \sum_{t=0}^{\tau_k^r-1} w_k^{r,t} \left\langle \nabla_k F(\bm\Theta^{r,0}), \bm g_k^{r,t} \right\rangle 
    \nonumber \\ &~~~~~~~~~~~~~~~~~~~
    + \frac{L}{2} \sum_{k=0}^K (\eta_k^r)^2 \tau_k^r \sum_{t=0}^{\tau_k^r-1} (w_k^{r,t})^2 \lrVert{\bm g_k^{r,t}}^2. 
    \label{beforeaddsub.eq}
\end{align}
where (\ref{beforeaddsub.eq}) follows from the fact that 
$(\sum_{n=1}^N \x_n)^2 \leq N\sum_{n=1}^N \x_n^2$.

We bound the first term in (\ref{beforeaddsub.eq}):
\begin{align}
    &-\left\langle \nabla_k F(\bm\Theta^{r,0}), \bm g_k^{r,t} \right\rangle \nonumber \\
    &=  
    -\left(\left\langle \nabla_k F(\bm\Theta^{r,0}), \bm g_k^{r,t} - \bm g_k^{r,0} \right\rangle 
    + \left\langle \nabla_k F(\bm\Theta^{r,0}), \bm g_k^{r,0} \right\rangle \right) \\
    &\leq  
    \frac{1}{2}\lrVert{\nabla_k F(\bm\Theta^{r,0})}^2 
        + \frac{1}{2}\lrVert{\bm g_k^{r,t} - \bm g_k^{r,0}}^2  
    \nonumber \\ &~~~~~~~~~~~~~~~~~~~~~~~~~~~~~~~~~~~~~
    - \left\langle \nabla_k F(\bm\Theta^{r,0}), \bm g_k^{r,0} \right\rangle 
    \label{product_half.eq}
\end{align}
where (\ref{product_half.eq}) follows from the fact that 
$A \cdot B = \frac{1}{2}A^2 + \frac{1}{2}B^2 - \frac{1}{2}(A-B)^2 
\leq \frac{1}{2}A^2 + \frac{1}{2}B^2$.

We bound the second term in (\ref{beforeaddsub.eq}):
\begin{align}
    \lrVert{\bm g_k^{r,t}}^2 
    &= \lrVert{\bm g_k^{r,t} - \bm g_k^{r,0} + \bm g_k^{r,0}}^2 \\
    &\leq  
    2\left(\lrVert{\bm g_k^{r,t} - \bm g_k^{r,0}}^2 + \lrVert{\bm g_k^{r,0}}^2 \right). 
    \label{normbound.eq}
\end{align}

Plugging (\ref{product_half.eq}) and (\ref{normbound.eq}) into (\ref{beforeaddsub.eq}), and 
applying the expectation $\mathbb{E}^r$ to both sides:
\begin{align}
    &\Ebatch{F(\bm\Theta^{r+1,0})} - F(\bm\Theta^{r,0}) \nonumber \\
    &\leq -\frac{1}{2}\sum_{k=0}^K \eta_k^r \sum_{t=0}^{\tau_k^r-1} w_k^{r,t} 
    (1-2\eta_k^r L \tau_k^r w_k^{r,t})\lrVert{\nabla_k F(\bm\Theta^{r,0})}^2 
    \nonumber \\ &~~~
    + \frac{1}{2}\sum_{k=0}^K \eta_k^r \sum_{t=0}^{\tau_k^r-1} w_k^{r,t}\Ebatch{\lrVert{\bm g_k^{r,t} - \bm g_k^{r,0}}^2} 
    \nonumber \\ &~~~
    + L \sum_{k=0}^K (\eta_k^r)^2 \tau_k^r \sum_{t=0}^{\tau_k^r-1} (w_k^{r,t})^2 
    \left(\Ebatch{\lrVert{\bm g_k^{r,t} - \bm g_k^{r,0}}^2} 
    + \sigma_k^2 \right) \\
    &= -\frac{1}{2}\sum_{k=0}^K \eta_k^r \sum_{t=0}^{\tau_k^r-1} w_k^{r,t} 
    (1-2\eta_k^r L \tau_k^r w_k^{r,t})\lrVert{\nabla_k F(\bm\Theta^{r,0})}^2 
    \nonumber \\ &~~~
    + \frac{1}{2} \sum_{k=0}^K \eta_k^r\sum_{t=0}^{\tau_k^r-1} w_k^{r,t} 
    (1+2\eta_k^r L \tau_k^r w_k^{r,t})\Ebatch{\lrVert{\bm g_k^{r,t} - \bm g_k^{r,0}}^2}
    \nonumber \\ &~~~
    + L \sum_{k=0}^K (\eta_k^r)^2 \tau_k^r \sum_{t=0}^{\tau_k^r-1} (w_k^{r,t})^2 \sigma_k^2. 
    \label{var1.eq}
\end{align}
Applying Lemma~\ref{diff.lemma} to (\ref{var1.eq}):
\begin{align}
    &\Ebatch{F(\bm\Theta^{r+1,0})} - F(\bm\Theta^{r,0}) \nonumber \\
    &\leq -\frac{1}{2}\sum_{k=0}^K \eta_k^r\sum_{t=0}^{\tau_k^r-1} w_k^{r,t} 
    (1-2\eta_k^r L \tau_k^r w_k^{r,t})\lrVert{\nabla_k F(\bm\Theta^{r,0})}^2 
    \nonumber \\ &~~~
    + 4\sum_{k=0}^K (\tau_k^r)^3(\eta_k^r)^2 L_k^2 
    \max_{t} (w_k^{r,t})^3 
    \left(\lrVert{\nabla_k F(\bm\Theta^{r,0})}^2 + \sigma_k^2\right) 
    \nonumber \\ &~~~
    + 8L\sum_{k=0}^K (\tau_k^r)^4(\eta_k^r)^3 L_k^2 
    \max_{t} (w_k^{r,t})^4 
    \left(\lrVert{\nabla_k F(\bm\Theta^{r,0})}^2 + \sigma_k^2\right) 
    \nonumber \\ &~~~
    + (\eta_k^r)^2 L \sum_{k=0}^K \tau_k^r \sum_{t=0}^{\tau_k^r-1} (w_k^{r,t})^2 \sigma_k^2. 
\end{align}

Let $\eta_k^r \leq \frac{1}{16 \tau_k^r \max\{L, L_k\} \max_{0 \leq t \leq \tau_k^r-1} w_k^{r,t}}$. 
Noticing that $\tau_k^r \leq \sum_{t=0}^{\tau_k^r-1} w_k^{r,t}$, we have:
\begin{align}
    &\Ebatch{F(\bm\Theta^{r+1,0})} - F(\bm\Theta^{r,0}) \nonumber \\
    &\leq -\frac{1}{2}\sum_{k=0}^K \eta_k^r\sum_{t=0}^{\tau_k^r-1} w_k^{r,t}
    \left(1-\frac{1}{8}
        -\frac{1}{32}
    -\frac{1}{256} \right)
    \lrVert{\nabla_k F(\bm\Theta^{r,0})}^2 
    \nonumber \\ &~~~
    + L 
    \sum_{k=0}^K (\eta_k^r)^2 \max_{0 \leq t \leq \tau_k^r-1} w_k^{r,t} \tau_k^r \sigma_k^2 
    \sum_{t=0}^{\tau_k^r-1} w_k^{r,t} \\
    &\leq -\frac{1}{4}\sum_{k=0}^K \eta_k^r\sum_{t=0}^{\tau_k^r-1} w_k^{r,t}
    \lrVert{\nabla_k F(\bm\Theta^{r,0})}^2 
    \nonumber \\ &~~~
    + L 
    \sum_{k=0}^K (\eta_k^r)^2 \max_{0 \leq t \leq \tau_k^r-1} w_k^{r,t} \tau_k^r \sigma_k^2 
    \sum_{t=0}^{\tau_k^r-1} w_k^{r,t}. 
\end{align}

After some rearranging of terms:
\begin{align}
    \sum_{k=0}^K \eta_k^r&\sum_{t=0}^{\tau_k^r-1} w_k^{r,t} 
    \lrVert{\nabla_k F(\bm\Theta^{r,0})}^2  \nonumber \\
    &\leq 4(F(\bm\Theta^{r,0}) - \Ebatch{F(\bm\Theta^{r+1,0})})  
    \nonumber \\ &~~~
    + 4L 
    \sum_{k=0}^K (\eta_k^r)^2 \max_{0 \leq t \leq \tau_k^r-1} w_k^{r,t} \tau_k^r \sigma_k^2 
    \sum_{t=0}^{\tau_k^r-1} w_k^{r,t}.
\end{align}

Next, we average over all global rounds $r=0,\ldots,R-1$ and take total expectation:
\begin{align}
     \frac{1}{R} \sum_{r=0}^{R-1} &\sum_{k=0}^K \eta_k^r\sum_{t=0}^{\tau_k^r-1} w_k^{r,t}
    \Etot{\lrVert{\nabla_k F(\bm\Theta^{r,0})}^2} \nonumber \\
    &\leq \frac{4(F(\bm\Theta^{0,0}) - \Etot{F(\bm\Theta^{R,0})})}{R} 
    \nonumber \\ &~~~
    + \frac{4L}{R} \sum_{r=0}^{R-1}  
    \sum_{k=0}^K (\eta_k^r)^2 \max_{0 \leq t \leq \tau_k^r-1} w_k^{r,t} \tau_k^r \sigma_k^2 
    \sum_{t=0}^{\tau_k^r-1} w_k^{r,t}.
\end{align}

In order to get our weighted averaged on the left-hand side, 
we divide through by $\sum_{r=0}^{R-1} \sum_{k=0}^K \eta_k^r\sum_{t=0}^{\tau_k^r-1} w_k^{r,t}$,
which completes the proof of Theorem~\ref{main.thm}.

\addtolength{\textheight}{-0.1cm}   %

\bibliography{references}
\bibliographystyle{IEEEtran}

\vspace{-10mm}
\begin{IEEEbiography}[{\includegraphics[width=1in,height=1.25in,clip,keepaspectratio]{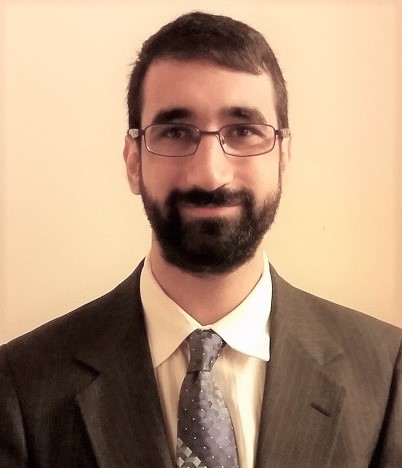}}]{Timothy Castiglia}
    is a field consultant at Ab Initio Software. 
    He received his Ph.D. degree from the Department
    of Computer Science at Rensselaer Polytechnic
    Institute, Troy, NY, USA. He received his
    B.S. degree in Computer Science in 2017 from the
    same institute. 
    His research interests include distributed systems,
    machine learning, and network science.
\end{IEEEbiography}
\vspace{-10mm}

\begin{IEEEbiography}[{\includegraphics[width=1in,height=1.25in,clip,keepaspectratio]{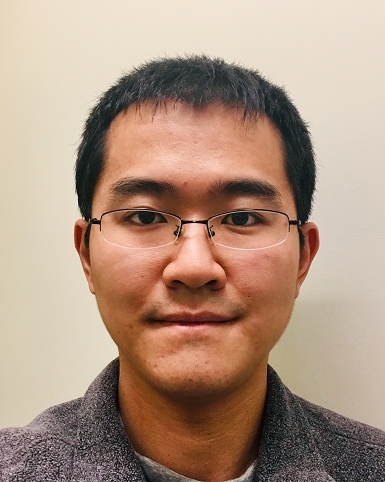}}]{Shiqiang Wang}
    (S’13–M’15–SM’23) received his Ph.D. from the Department of Electrical and Electronic Engineering, Imperial College London, United Kingdom, in 2015. He is currently a Staff Research Scientist at IBM T. J. Watson Research Center, NY, USA. His current research focuses on the intersection of distributed computing, machine learning, networking, and optimization, with a broad range of applications including data analytics, edge-based artificial intelligence (Edge AI), Internet of Things (IoT), and future wireless systems. He has made foundational contributions to edge computing and federated learning that generated both academic and industrial impact. Dr. Wang serves as an associate editor of the IEEE Transactions on Mobile Computing and IEEE Transactions on Parallel and Distributed Systems. He received the IEEE Communications Society (ComSoc) Leonard G. Abraham Prize in 2021, IEEE ComSoc Best Young Professional Award in Industry in 2021, IBM Outstanding Technical Achievement Awards (OTAA) in 2019, 2021, 2022, and 2023, multiple Invention Achievement Awards from IBM since 2016, Best Paper Finalist of the IEEE International Conference on Image Processing (ICIP) 2019, and Best Student Paper Award of the Network and Information Sciences International Technology Alliance (NIS-ITA) in 2015.
\end{IEEEbiography}
\vspace{-10mm}

\begin{IEEEbiography}[{\includegraphics[width=1in,height=1.25in,clip,keepaspectratio]{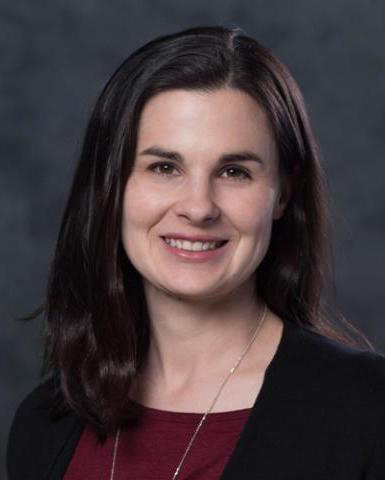}}]{Stacy Patterson}
    is an Associate Professor in the Department
    of Computer Science at Rensselaer Polytechnic
    Institute. She received the MS and PhD in
    computer science from UC Santa Barbara in 2003
    and 2009, respectively. From 2009-2011, she was
    a postdoctoral scholar at the Center for Control,
    Dynamical Systems and Computation at UC Santa
    Barbara. From 2011-2013, she was a postdoctoral
    fellow in the Department of Electrical Engineering
    at Technion - Israel Institute of Technology. Dr.
    Patterson is the recipient of a Viterbi postdoctoral
    fellowship, the IEEE CSS Axelby Outstanding Paper Award, and an NSF
    CAREER award. Her research interests include distributed systems, cloud
    computing, and the Internet of Things.
\end{IEEEbiography}

\end{document}